\documentclass[twocolumn,notitlepage,prl,superscriptaddress,longbibliography]{revtex4-2}
\usepackage{amsfonts}
\usepackage{textcomp}
\usepackage{times}
\usepackage{graphicx}
\usepackage{float}
\usepackage{latexsym,amsmath,amssymb,bm,euscript}
\usepackage{color}
\usepackage{subfigure}
\usepackage{epstopdf}
\usepackage[colorlinks,
linkcolor=blue,
anchorcolor=blue,
citecolor=blue,
urlcolor=blue,]{hyperref}
\usepackage{hyperref}
\usepackage{soul}
\usepackage[normalem]{ulem}
\usepackage{mathrsfs}
\usepackage{amsmath}
\usepackage{xspace}
\usepackage{natbib}
\usepackage{ulem}
\usepackage{etoolbox}
\usepackage{tikz}
\usepackage{array}

\begin{document}
	%	\title{Metal-insulator Transitions and Structural Distortion Originated from Charge Density Wave and Oxygen Vacancies in Superconducting La$ _{3} $Ni$ _{2} $O$ _{7} $}
	\title{Antiferromagnetic Ground State, Charge Density Waves and Oxygen Vacancies Induced Metal-Insulator Transition in Pressurized La$ _{3} $Ni$ _{2} $O$ _{7} $}
	\author{Xin-Wei Yi}
	\affiliation{School of Physical Sciences, University of Chinese Academy of Sciences, Beijing 100049, China}
	
	\author{Ying Meng}
	\affiliation{School of Physical Sciences, University of Chinese Academy of Sciences, Beijing 100049, China} \affiliation{Beijing National Laboratory for Condensed Matter Physics, Institute of Physics, Chinese Academy of Sciences, Beijing 100190, China.}
	
	\author{Jia-Wen Li}
	\affiliation{School of Physical Sciences, University of Chinese Academy of Sciences, Beijing 100049, China} \affiliation{Kavli Institute for Theoretical Sciences, University of Chinese Academy of Sciences, Beijing 100190, China}
	
	\author{Zheng-Wei Liao}
	\affiliation{School of Physical Sciences, University of Chinese Academy of Sciences, Beijing 100049, China}
	
	\author{Jing-Yang You}
	\email{phyjyy@nus.edu.sg}
	\affiliation{Department of Physics, Faculty of Science, National University of Singapore, 117551, Singapore}
	
	\author{Bo Gu}
	\email{gubo@ucas.ac.cn}	
	\affiliation{School of Physical Sciences, University of Chinese Academy of Sciences, Beijing 100049, China} \affiliation{Kavli Institute for Theoretical Sciences, University of Chinese Academy of Sciences, Beijing 100190, China}
	
	\author{Gang Su}
	\email{gsu@ucas.ac.cn}
	\affiliation{School of Physical Sciences, University of Chinese Academy of Sciences, Beijing 100049, China} \affiliation{Kavli Institute for Theoretical Sciences, University of Chinese Academy of Sciences, Beijing 100190, China}
	
	\begin{abstract}
		La$ _{3} $Ni$ _{2} $O$ _{7} $ has garnered widespread interest recently due to its high-temperature superconductivity under pressure, accompanied by charge density wave (CDW) ordering and metal-insulator (MI) transitions in the phase diagram. Here, we reveal with comprehensive calculations that La$ _{3} $Ni$ _{2} $O$ _{7} $ possesses an antiferromagnetic ground state under both low and high pressures, with the strong Fermi surface nesting contributed by the flat band that leads to phonon softening and electronic instabilities. Several stable CDW orders with oxygen octahedral distortions are identified, which can trigger the MI transitions. The estimated CDW transition temperature ($\approx$120 K) at ambient pressure agrees nicely with experimental results. In the presence of apical oxygen vacancies, we identify two different phases, say, half distortion and full distortion phases, respectively, and their competition can lead to a pressure-induced MI transition, in good agreement with experimental observations. In addition, we find that the electron-phonon coupling is too small to contribute to superconductivity. These results appear to indicate an unconventional superconducting pairing mechanism mediated by antiferromagnetic fluctuations. A phase diagram that is consistent with the experimental results is given. The present results not only explain the origins of experimentally observed CDW and MI transitions, but also provide insight for deeply understanding the properties like superconductivity, CDW and the role of oxygen vacancies in pressurized La$ _{3} $Ni$ _{2} $O$ _{7} $.

		%\begin{description}
		%	\item[keywords]
		%1 2 3
		%\end{description}
	\end{abstract}
	
	\maketitle
	%\tableofcontents
	
	\textit{Introduction}.---Over the past decades, exploring and understanding superconductors with high superconducting transition temperatures ($\textit{T}_\textit{c}$) have been the pivotal focuses in condensed matter physics \cite{RN902, RN1535}. Recently, the Ruddlesden–Popper nickelate La$ _{3} $Ni$ _{2} $O$ _{7} $ was found to exhibit a $\textit{T}_\textit{c}$ of nearly 80 K under pressure \cite{RN1300}. Subsequent experiments consistently confirmed this observation, as evidenced by both zero resistance \cite{2307.14819, RN1305, RN1515, RN1417, 2311.12361} and diamagnetic susceptibility \cite{2311.12361}. The electronic configuration with Ni$ ^{2.5+} $ in La$ _{3} $Ni$ _{2} $O$ _{7} $ deviates from the 3$\textit{d}^\textit{9} $ configuration of transition metal ions in infinite-layer nickelates and cuprates \cite{RN1300,RN1389, RN1308}. Despite diverse views in extensive theoretical studies, broad consensuses have emerged regarding the crucial roles played by interlayer antiferromagnetic coupling, hybridization of $\textit{e}_\textit{g} $ orbitals, Hubbard interaction, Hund coupling, etc., for the onset of superconductivity \cite{RN1400, RN1399, RN1323, RN1321, 2306.06039, 2306.07275, RN1324, RN1326, RN1350, RN1329, RN1331,2307.14965, RN1334, 2307.16697, RN1451, RN1337, RN1397, 2308.09698,2308.11614, 2308.11195, 2308.16564, 2309.05726, 2309.15095, 2309.17279, 2310.02915, RN1430, 2310.17465, 2311.03349, 2311.05491, 2311.07316, 2311.09970, 2311.10066, 2311.12769, 2312.03605, 2312.04304, 2312.04401, 2312.17064, RN1553}.
	
	Apart from superconductivity, the system undergoes a potential charge density wave (CDW) transition around 110-140 K under low pressures, as evidenced by nuclear magnetic resonance and anomalous kinks of resistivity, magnetization, specific heat and optical conductivity \cite{ RN1353, 2312.11844, RN1346, RN1356, RN1348, RN1305, 2307.02950, 2307.14819, RN1417}. On the other hand, despite some La$ _{3} $Ni$ _{2} $O$ _{7} $ samples being identified as metals under ambient pressure, they undergo an unexpected metal-insulator (MI) transition under 2-4 GPa \cite{ RN1300}. In contrast, the calculations based on the non-magnetic (NM) state reveal a robust metallic character under both low and high pressures \cite{ RN1300, 2307.15276, 2309.17279,2309.01148}. Therefore, an integrated and reasonable explanation for these experimental results remains elusive so far.
	
	In this Letter, we perform comprehensive calculations based on the density functional theory (DFT) to examine structural, electronic, and phonon properties of La$ _{3} $Ni$ _{2} $O$ _{7} $ under pressures. We find that La$ _{3} $Ni$ _{2} $O$ _{7} $ possesses an antiferromagnetic ground state under both low and high pressures, with the strong Fermi surface (FS) nesting of electronic flat band that gives rise to the phonon softening at high symmetric points and electronic instabilities. We elucidate that CDW orders are originated from Peierls instability and trigger MI transitions. It is also found that the competition between the half distortion and full distortion phases with apical oxygen vacancies induces an MI transition under pressures. These calculated results are in good agreement with recent experimental findings. In addition, the electron-phonon coupling (EPC) is uncovered insufficient to be responsible for superconductivity, suggesting that the Cooper pairing mechanism is unconventional in pressurized La$ _{3} $Ni$ _{2} $O$ _{7} $, and is probably from antiferromagnetic fluctuations. 
	
\begin{figure}[t]
		\centering
		\includegraphics[scale=0.535,angle=0]{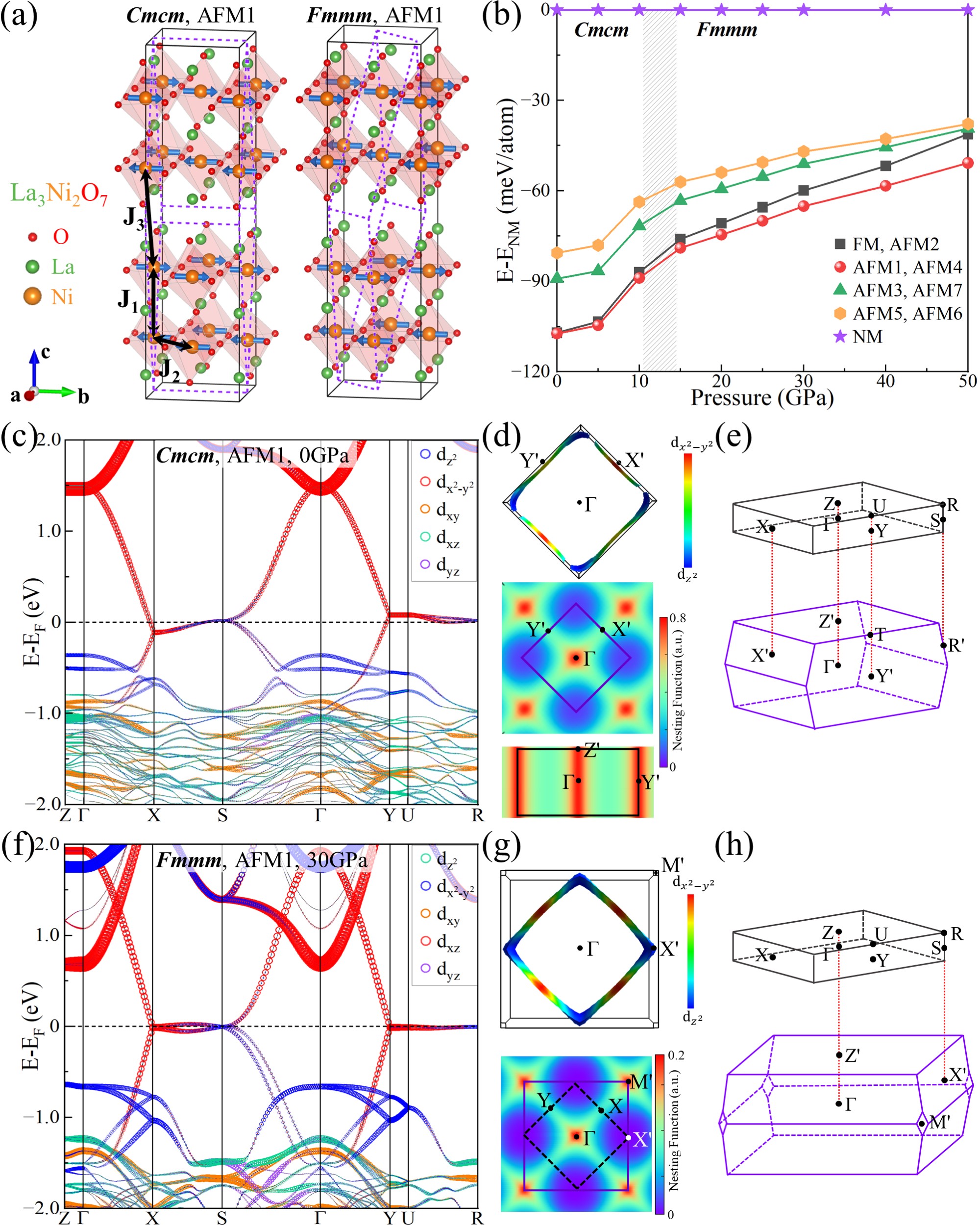}\\
		\caption{(a) Crystal structures for the $ \textit{Cmcm} $ and $ \textit{Fmmm} $ phases of La$ _{3} $Ni$ _{2} $O$ _{7} $. Black solid and purple dotted lines represent the conventional and primitive cells, respectively. Blue arrows indicate magnetic moment orientations for AFM1 ground state. $\textit{J}_\textit{1}$, $\textit{J}_\textit{2}$ and $\textit{J}_\textit{3}$ represent the interlayer nearest-neighbor (NN) coupling, intralayer NN coupling and interlayer next NN coupling among Ni atoms, respectively. (b) Relative energies of eight magnetic configurations with respect to the NM state under pressure. (c) Band structure projected by Ni orbitals of the $ \textit{Cmcm} $ phase under AFM1 configuration at 0 GPa. (d) Fermi surface projected by Ni-$d_{x^{2}-y^{2}}$ and $d_{z^{2}}$ orbitals and its nesting function at 0 GPa. (e) Brillouin zone (BZ) of Bravais and primitive lattices depicted by black and purple lines, respectively. (f-h) Same as (c-e) but for the $ \textit{Fmmm} $ phase at 30 GPa.}\label{band}
	\end{figure}

	\textit{Superconductivity not from BCS}.---The crystal structure of La$ _{3} $Ni$ _{2} $O$ _{7} $ changes from the $ \textit{Cmcm} $ to the $ \textit{Fmmm} $ space group at about 15 GPa [FiG.~\ref{band}(a)] \cite{RN1300}, coinciding with the onset of superconductivity. Based on Bardeen-Cooper-Schrieffer (BCS) scenario, we calculate the superconducting properties of La$ _{3} $Ni$ _{2} $O$ _{7} $ as listed in Table~\ref{table1}. Despite a considerable electronic density of states at the Fermi level N($\textit{E}_\textit{F}$), we find that the EPC constant $\lambda$ remains below 0.25 for both $ \textit{Cmcm} $ and $ \textit{Fmmm} $ phases, and both N($\textit{E}_\textit{F}$) and $\lambda$ decrease with increasing pressure. By calculations we reveal that the low $\lambda$ results in zero $\textit{T}_\textit{c}$ under 0-50 GPa, indicating that the EPC alone cannot account for experimental $\textit{T}_\textit{c, exp.}$ \cite{RN1300}. This shows that the superconducting pairing mechanism in pressurized La$ _{3} $Ni$ _{2} $O$ _{7} $ is unconventional.
	
	\begin{table}
		%\begin{table*}\footnotesize\small\normalsize
		%\renewcommand\arraystretch{1.15}
		\caption{Electron-phonon coupling constant $\lambda$, electronic density of states per formula unit (f.u.) at Fermi energy N($\textit{E}_\textit{F}$), logarithmic average frequency $\omega_\textit{log}$, estimated $\textit{T}_{\textit{c, BCS}}$ and $\textit{T}_{\textit{c, exp.}}$ \cite{RN1300} under pressure.}
		{\centering
			\begin{tabular}{lp{1cm}<{\centering}p{1.2cm}<{\centering}p{0.75cm}<{\centering}p{1.75cm}<{\centering}p{0.95cm}<{\centering}p{1cm}<{\centering}p{0.8cm}<{\centering}p{0.8cm}}
				\hline
				\hline
				&Pressure (GPa) &Space group &$\lambda$ &N($\textit{E}_\textit{F}$) (eV$^{-1}$f.u.$^{-1}$) &$\omega_\textit{log}$ (K) &$\textit{T}_{\textit{c, BCS}}$ (K) &$\textit{T}_{\textit{c, exp.}}$\\
				\hline
				&0 & $ \textit{Cmcm} $ &0.245 &43.50 &129.17 &0.01 &0\\
				&10 & $ \textit{Cmcm} $ &0.166 &37.96 &178.63 &0.00 &0\\
				&20 & $ \textit{Fmmm} $ &0.155 &37.93 &304.89 &0.00 &82\\
				&30 & $ \textit{Fmmm} $ &0.145 &36.25 &306.15 &0.00 &75\\
				&40 & $ \textit{Fmmm} $ &0.140 &35.39 &312.92 &0.00 &72\\
				&50 & $ \textit{Fmmm} $ &0.137 &34.68 &324.51 &0.00 &-\\
				\hline
				\hline	
		\end{tabular}}\label{table1}
	\end{table}

	\textit{Antiferromagnetic ground state}.---We consider the interlayer nearest-neighbor (NN) coupling ($\textit{J}_\textit{1}$), intralayer NN coupling ($\textit{J}_\textit{2}$), and interlayer next NN coupling ($\textit{J}_\textit{3}$) among Ni atoms [FiG.~\ref{band}(a)]. Combining all positive and negative values of $\textit{J}_\textit{1}$, $\textit{J}_\textit{2}$ and $\textit{J}_\textit{3}$, we calculate ferromagnetic (FM) and seven antiferromagnetic (AFM1-7) configurations, as depicted in FIG. S1 of supplemental materials (SM) \cite{supp}. As seen in FIG.~\ref{band}(b), the energies of eight magnetic configurations form degenerate pairs, originating from $J _{3} \approx $ 0. Energies of all magnetic states are lower than that of NM state, suggesting magnetic tendencies favor the stability of system. The magnetic ground state is AFM1 (AFM4), which shows intralayer ferromagnetism and interlayer antiferromagnetism [FiG.~\ref{band}(a)]. Notably, experimental results including muon spin relaxation ($\mu$SR), resonant inelastic X-ray scattering and magnetic susceptibility also support the AFM ground state \cite{2311.15717, 2402.10485, 2401.12657, RN1356}. The AFM ground state suggests that the Cooper pairing in this nickelate may be mediated by AFM fluctuations.
	
	\textit{Fermi surface nesting}.---Under the AFM1 configuration, the band structures of the $ \textit{Cmcm} $ phase at 0 GPa and the $ \textit{Fmmm} $ phase at 30 GPa are depicted in FIG.~\ref{band}(c) and (f), respectively. Similar to the NM scenario, their electronic states near $\textit{E}_\textit{F}$ are predominantly contributed by the $\textit{d}_{x^{2}-y^{2}} $ and $\textit{d}_{z^{2}} $ orbitals. However, a remarkably flat band emerges near $\textit{E}_\textit{F}$, with its flatness significantly increasing as pressure rises to 30 GPa, suggesting an enhanced correlated electron effect. More intriguingly, the FS formed by this flat band exhibits a perfect two-dimensional (2D) square-cylinder characteristic in FIG.~\ref{band}(d) and (g). The band characteristics near $\textit{E}_\textit{F}$ in La$ _{3} $Ni$ _{2} $O$ _{7} $ resemble the results of the Hubbard model on a 2D square lattice \cite{RN1483}. The strong nesting features of this square FS evoke potential electronic instabilities. The calculated FS nesting function $\xi(q)$ in FIG.~\ref{band}(d) \cite{RN1480, supp}, shows strong peaks along $\Gamma$-Z$^{\prime}$, which aligns with the nesting vector of the FS. Results at 30 GPa in FIG.~\ref{band}(g) is similar, albeit with a different BZ and high symmetry points.
	
	\begin{figure*}[t]
		\centering
		\includegraphics[scale=0.44,angle=0]{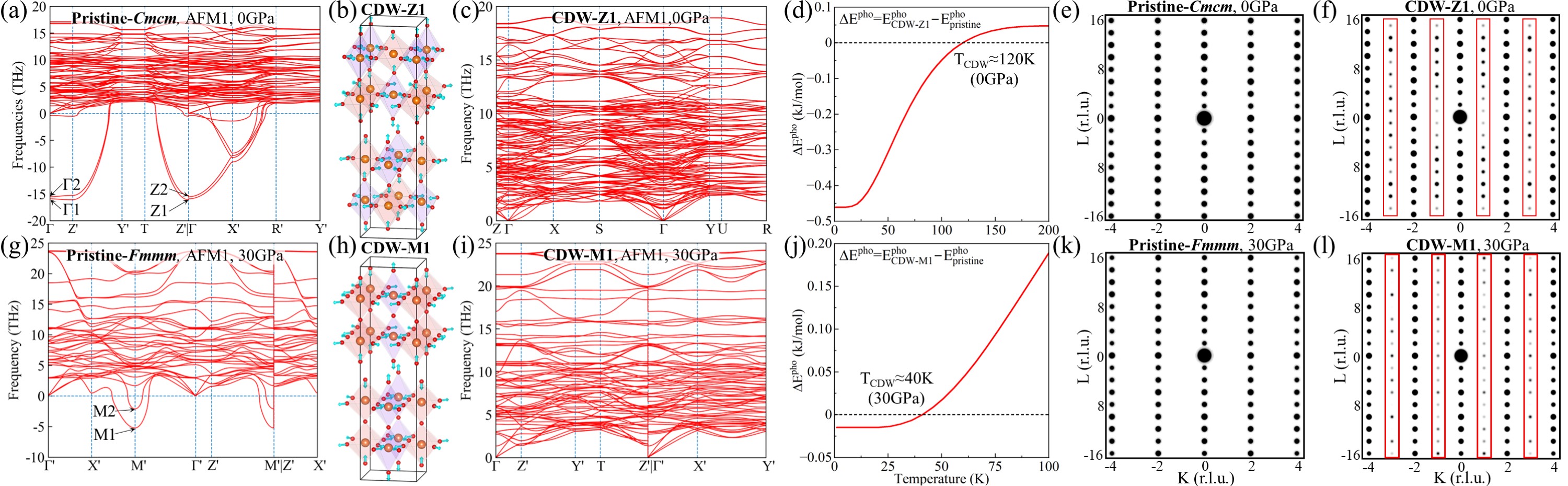}\\
		\caption{(a) The phonon spectrum of $ \textit{Cmcm} $ phase at 0 GPa under AFM1 configuration. (b) Crystal structure and (c) phonon spectrum of CDW-Z1 phase corresponding to imaginary mode Z1. Oxygen octahedra with in-plane expansion (fat) and contraction (skim) are depicted with light red and purple colors, respectively. (d) The difference in harmonic phonon energy between the AFM1 state of CDW-Z1 phase and the NM state of $ \textit{Cmcm} $ phase. Simulated diffraction patterns of (e) $ \textit{Cmcm} $ and (f) CDW-Z1 phases along the [100] zone axis. Miller indices $ \textit{K} $ and $ \textit{L} $ are given in reciprocal lattice unit (r.l.u.). Emergent Bragg peaks of CDW phase comparing with pristine phase are highlighted by red boxes. (g-l) The same as (a-f) but for the $ \textit{Fmmm} $ phase and CDW-M1 phase at 30 GPa.}\label{trans}
	\end{figure*}

	\textit{Stable CDW phases}.---To explore its structural stability, the phonon spectra at 0 and 30 GPa are calculated in FIG.~\ref{trans}(a) and (g). For the $ \textit{Cmcm} $ phase, two phonon branches exhibit imaginary frequencies along the $\Gamma$-Z$ ^{\prime} $ path, with four imaginary modes at $\Gamma$ and Z$ ^{\prime} $ points denoted as $\Gamma1$, $\Gamma2$, Z1, and Z2, respectively, each leading to distinct structural distortions. Following careful structural relaxation, stable phases are obtained named as CDW-$\Gamma1$ (space group: $\textit{P2}_{1}\textit{/m}$), CDW-$\Gamma2$ ($\textit{Amm2}$), CDW-Z1 ($\textit{Pnma}$) and CDW-Z2 ($\textit{Pmmn}$), respectively. Detailed lattice information for these phases can be found in Table S3. These phases share a similarity in the in-plane expansion (fat) and contraction (skim) of oxygen octahedra alternating in each layer, with difference lying in the arrangement and orientation of the octahedra [FIG.~\ref{trans}(b) and S5]. For the CDW-Z1 phase corresponding to the lowest imaginary mode Z1 as an example, the phonon spectrum in FIG.~\ref{trans}(c) confirms its stability. By comparing the harmonic phonon energy of the NM state of pristine phase with the AFM1 state of the CDW-Z1 phase, we estimate the transition temperature of CDW phase ($\textit{T}_\textit{CDW}$) to be $\approx$120 K [FIG.~\ref{trans}(d)], consistent with experimental results \cite{ RN1353, RN1346, RN1356, RN1348, RN1305, 2307.02950, 2307.14819, RN1417, 2312.11844}.
	
	For the $ \textit{Fmmm} $ phase, only phonon branches near the M$ ^{\prime} $ point show imaginary frequencies, with a softer effect compared to the $ \textit{Cmcm} $ phase at 0 GPa [FiG.~\ref{trans}(g)]. Two imaginary modes at the M$ ^{\prime} $ point are labeled as M1 and M2, corresponding to stable distorted structures named CDW-M1 ($\textit{Cmmm}$) and CDW-M2 ($\textit{Cmcm}$), respectively. CDW-M1 exhibits similar octahedral distortions to CDW-Z1 but includes an additional horizontal mirror symmetry of the Ni-O bilayer [FiG.~\ref{trans}(h)]. The estimated $\textit{T}_\textit{CDW}$ of CDW-M1 at 30 GPa is around 40 K, much lower than that at 0 GPa. Other CDW phases at various pressures also exhibit stable phonon spectra, as shown in FIG. S5 and S6.
	
	Notably, the momenta with imaginary phonons in FIG.~\ref{trans}(a) and (g) correspond to those exhibiting strong peaks of $\xi(q)$ in FIG.~\ref{band}(d) and (g), confirming that CDW originates from Peierls instability related to FS nesting. The imaginary phonons in the AFM1 state contrasts with the stable phonons in the NM state \cite{2307.15276}, implying the instabilities are related to antiferromagnetism.
	
	The distortion of Ni-O bond length is less than 0.1 \AA, posing challenges in probing predicted CDW structures in experiments. Simulated XRD spectra show that most CDW phases exhibit nearly identical peak positions and intensities to pristine phases at both 0 and 30 GPa, except for CDW-$\Gamma$1 with a distinct Bravais lattice [FiG. S9 and S10]. To aid in identifying CDW phases in experiments, we simulate diffraction patterns that can be measured using techniques such as XRD and transmission electron microscopy. Diffraction patterns of the CDW-Z1 phase show emergent spots with odd Miller indices $ \textit{K} $ compared with pristine-$ \textit{Cmcm} $ phases in FIG.~\ref{trans}(e) and (f). Similar patterns are observed in the CDW-M1 phase at 30 GPa in FIG.~\ref{trans}(l) and other CDW phases in FIG. S11 and S12.
	
	\begin{figure}[b]
		\centering
		\includegraphics[scale=0.47,angle=0]{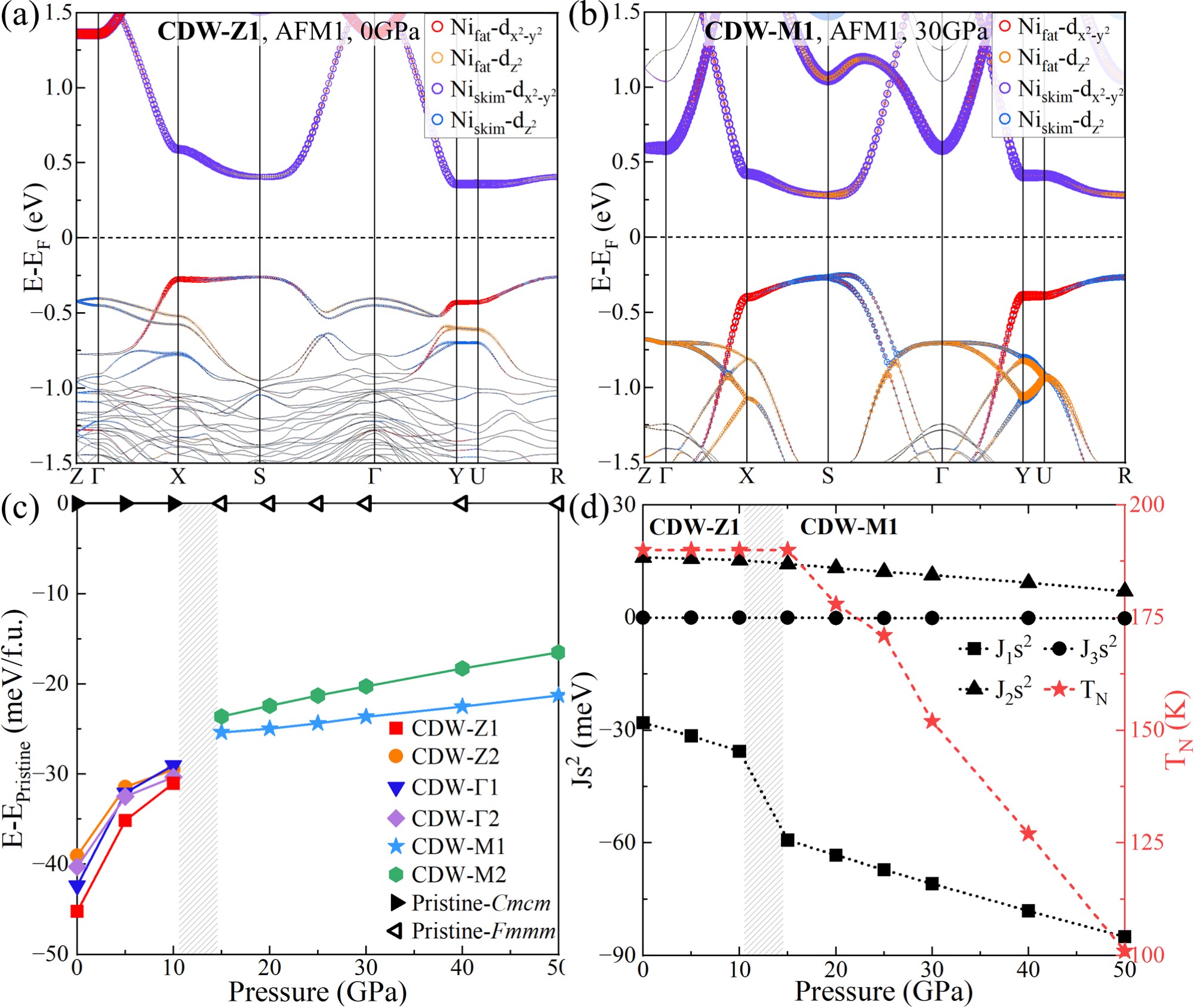}\\
		\caption{%Electronic and magnetic properties for CDW phases.
			Band structures for (a) CDW-Z1 phase at 0GPa and (b) CDW-M1 phases at 30GPa. Ni$ _\mathrm{fat(skim)} $ represents Ni atoms in fat (skim) oxygen oxtahedra. (c) Relative energies of CDW phases with respect to the pristine phases under pressure. (d) Interlayer NN exchange coupling ($\textit{J}_\textit{1}$), intralayer NN exchange coupling ($\textit{J}_\textit{2}$), and interlayer next NN exchange coupling ($\textit{J}_\textit{3}$) and Neel temperature $\textit{T}_\textit{N}$ under pressure for CDW-Z1 and CDW-M1 phases.}\label{presstopo}
	\end{figure}
	
	We further analyze the magnetic properties of CDW phases. By comparing the energy of different magnetic states under pressure [FiG. S1], we find a gradual increase in the AFM $\textit{J}_\textit{1}$, alongside a decrease in the FM $\textit{J}_\textit{2}$ with increasing pressure, except for $\textit{J}_\textit{3} \approx$ 0 as seen in Fig.~\ref{presstopo}(d). Furthermore, $|\textit{J}_\textit{1} |$ consistently exceeds $\textit{J}_\textit{2}$, indicating strong interlayer AFM coupling. Employing Monte Carlo simulation based on the Heisenberg-type Hamiltonian: $\textit{H =--}\sum_\textit{i}^\textit{1,2,3} \textit{J}_\textit{i}\cdot\textbf{s}^2\textit{+E}_\textit{0}$, where $\textit{E}_\textit{0}$ is the energy independent of the spin configurations, we determine that the Neel temperature $\textit{T}_\textit{N}$ remains above 100 K under 0-50 GPa, surpassing the calculated $\textit{T}_\textit{CDW}$ in FIG.~\ref{trans}(g) and (j), suggesting that the presence of robust magnetic coupling is capable of inducing the CDW transition.
	
	Unlike pristine phases with metallic behavior in FIG.~\ref{band}(c) and (f), both CDW-Z1 and CDW-M1 phases exhibit band gaps of about 0.5 eV in FIG.~\ref{presstopo}(a) and (b). Similar gaps are observed in other CDW phases [FIG. S5 and S6]. The orbitals of Ni atoms in the expanded and contracted oxygen octahedra (named as Ni$_\mathrm{fat}$ and Ni$_\mathrm{skim}$) are split, with the $\textit{d} _{x^{2}-y^{2}} $ orbital shifted downward and upward, respectively. It means that the distinct octahedra caused by CDW transitions leads to this MI transition.

	All CDW phases have lower enthalpies compared to pristine phases under pressure in FIG.~\ref{presstopo}(c), indicating their energetic stability. Specifically, CDW-Z1 and CDW-M1 originating from the lowest imaginary phonon modes have the lowest energy under low and high pressure, respectively.
	
	\begin{figure}[t]
		\centering
		\includegraphics[scale=0.47,angle=0]{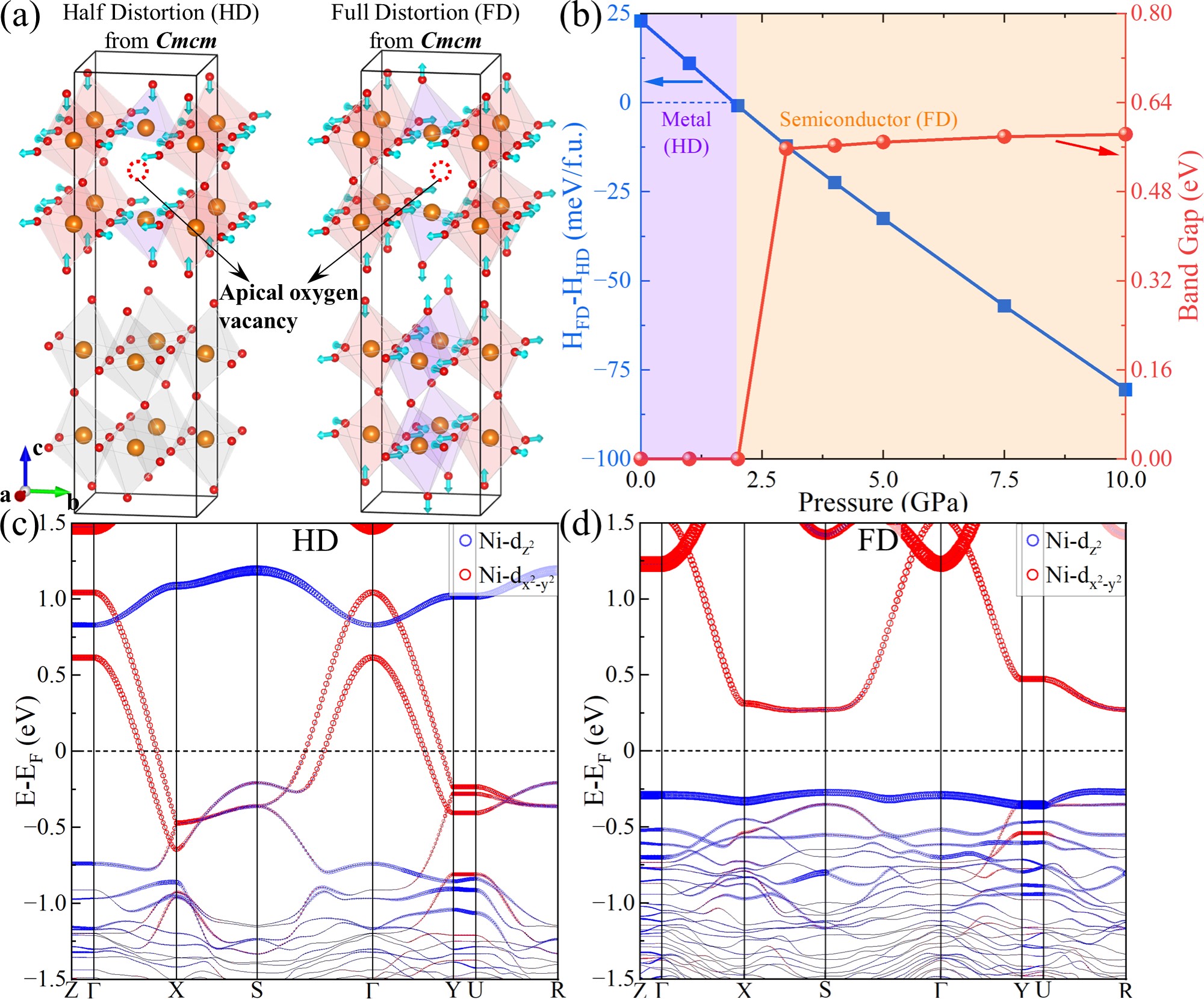}\\
		\caption{(a) Two phases of La$ _{3} $Ni$ _{2} $O$ _{6.75} $ optimized from pristine $ \textit{Cmcm} $ phase, labeled as half distortion (HD) and full distortion (FD). (b) Enthalpy (\textit{H}) difference between two phases and band gaps of the phase with lower \textit{H} under pressure. Band structures for (c) HD and (d) FD phases at 0 GPa.}\label{dope}
	\end{figure}
	
	\textit{Role of Apical Oxygen Vacancies}.---Experimental evidence suggests the existence of oxygen vacancies with a nonstoichiometry $\delta$ of La$ _{3} $Ni$ _{2} $O$_{7-\delta}$ as high as 0.34 \cite{2312.15727}. Our calculations reveal that the apical oxygen vacancy has the lowest energy compared with vacancies at other sites, consistent with previous reports \cite{2312.15727,2312.01271}. After careful structural optimization for La$ _{3} $Ni$ _{2} $O$_{6.75}$ with apical oxygen vacancies, two unexpectedly stable structures emerge as shown in FIG.~\ref{dope}(a). One structure exhibits structural distortion only near the defect, while the other shows distortion in all NiO bilayers, labeled as half distortion (HD) and full distortion (FD) phases, respectively. Comparing their enthalpies ($ \textit{H} $) in FIG.~\ref{dope}(b), we observe a phase transition from the HD to the FD phase at 2 GPa. The $ \textit{H} $ can be divided into two terms, $\textit{H=E+PV}$, where $\textit{E}$, $\textit{P}$, and $ \textit{V} $ represent energy, pressure, and volume, respectively. As seen in FIG. S13, although the HD phase consistently displays lower $E$ at 0-10 GPa, the smaller $ \textit{V} $ of the FD phase makes it thermodynamically more stable under high pressure.
	
	Further band calculations reveal the HD phase is a metal similar to electron doped La$ _{3} $Ni$ _{2} $O$ _{7} $ in FIG.~\ref{band}(c). Conversely, the FD phase exhibits a gap of about 0.5 eV and a flat valence band. This indicates an MI transition accompanying the phase transition of the two vacancy structures, which is absent in the pure crystal phase. The FD phase of La$ _{3} $Ni$ _{2} $O$ _{6.75} $ in FIG.~\ref{dope}(a) exhibits expansion and contraction for all oxygen octahedra, similar to the CDW phase in FIG.~\ref{trans}(h). This unexpected similarity is intriguing. The lower $\textit{H}$ of the FD phase above 2 GPa indicates harmony between oxygen vacancies and CDW under high pressure. On the contrary, the HD phase becomes more stable below 2 GPa, suggesting oxygen vacancies suppress structural distortions of CDW at this stage.
	
	\begin{figure}[t]
		\centering
		\includegraphics[scale=0.4,angle=0]{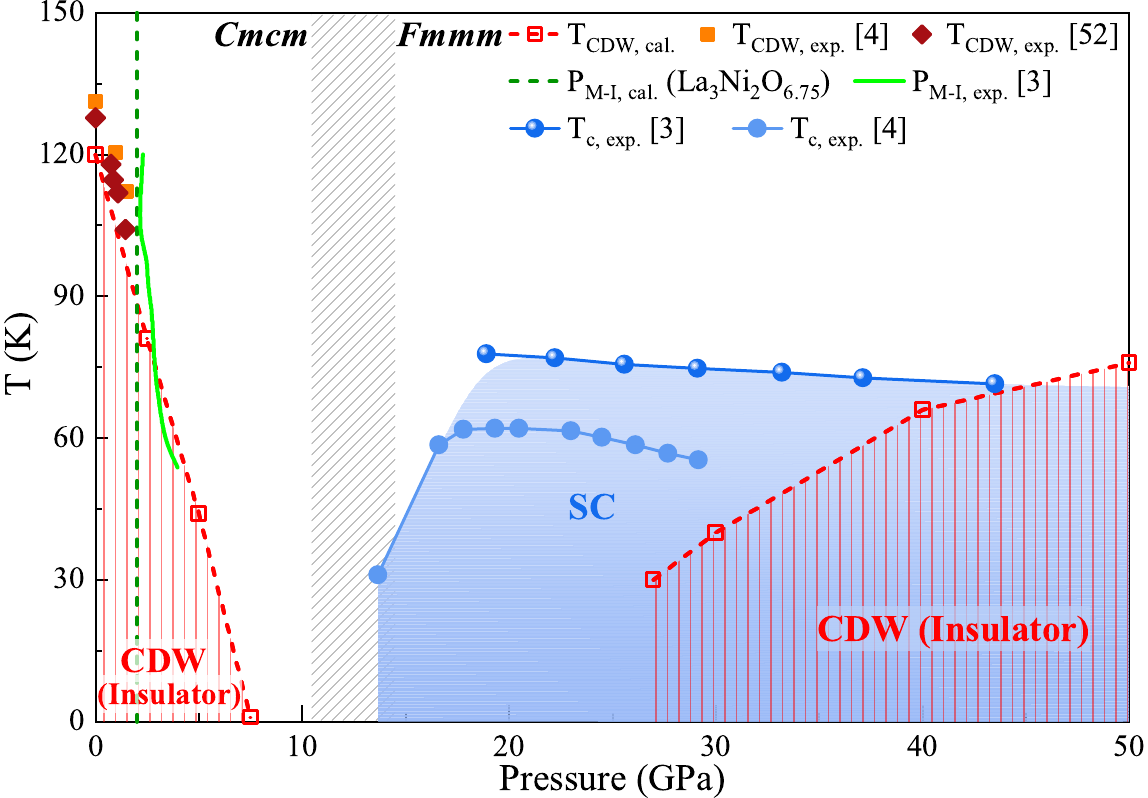}\\
		\caption{Phase diagram of La$ _{3} $Ni$ _{2} $O$ _{7} $. Solid lines and dots represent experimental results (exp.), while dashed lines and hollow dots represent calculated results (cal.) \cite{2307.14819,RN1356,RN1300,2307.14819}.}\label{phasediagram}
	\end{figure}
	
	\textit{Phase Diagram and Discussions}.---
	We summarize the main findings and related experimental data in a phase diagram shown in FIG.~\ref{phasediagram}. It can be seen that the calculated $\textit{T}_\textit{CDW, cal.}$ of the CDW-Z1 phase gradually decreases with increasing pressure, reaching zero at 7.5 GPa, nicely aligning with experimental results \cite{2307.14819, RN1356, RN1417}. Our calculations clearly reveal that the temperature-induced MI transition stems from the structural distortion of CDW \cite{RN1305, RN1515, RN1356, RN1366}. Beyond 27 GPa, the CDW-M1 phase emerges with $\textit{T}_\textit{CDW, cal.} > $ 30 K. Recent experiments demonstrate a pronounced insulating behavior under high pressure below 50 K compressed in the diamond anvil cell and above $\textit{T}_\textit{c}$ compressed in the cubic anvil cell \cite{RN1305}. These findings may be related to our calculated CDW phases under high pressure, suggesting a potential coexistence or competition between CDW and superconductivity.
	
	We found that the nonstoichiometric La$ _{3} $Ni$ _{2} $O$ _{6.75} $ undergoes an MI transition at 2 GPa, in good consistent with the pressure of MI transition in experiments ($\textit{P}_\textit{M-I, exp.}$) as seen in FIG.~\ref{phasediagram} \cite{RN1300}. The HD and FD phases identified in this work gain the insights for further theoretical and experimental analyses of the role of oxygen vacancies.
	
	The $\mu$SR experiments reveal a bulk AFM transition near 150 K for La$ _{3} $Ni$ _{2} $O$_{7-\delta}$ \cite{2311.15717,2402.10485}. Additionally, a continuous increase in magnetic susceptibility suggests a potential AFM order below 350 K for La$ _{3} $Ni$ _{2} $O$_{7.05}$ \cite{RN1356}. Our calculated $\textit{T}_\textit{N}$ for the CDW-Z1, pristine-$ \textit{Cmcm} $ and HD phases are 190, 393 and 95 K at 0 GPa, respectively [Table S2], showing that such variation in $\textit{T}_\textit{N, cal.}$ among different phases may explain the inconsistency in reported experimental $\textit{T}_\textit{N}$.
	
	%The properties of La$ _{3} $Ni$ _{2} $O$_{7-\delta}$, such as resistance, magnetization, CDW and superconductivity, are sensitively sample dependent \cite{ RN1348, RN1426, RN1346, RN1366, RN1305}. Our calculations reveal diverse structural phases, including multiple CDW and oxygen-vacancy phases. Understanding these intertwining properties requires exploring growth conditions and methods for different structural phases.
	
	%CDW and spin-density wave (SDW) can coexist in systems with strong electronic interactions. For instance, commensurate or incommensurate charge and spin modulations were observed in La$ _{4} $Ni$ _{3} $O$ _{8} $ \cite{RN1396, RN1564}, doped La$ _{2} $NiO$ _{4} $ \cite{RN1565, RN1566,RN1567}, La$ _{4} $Ni$ _{3} $O$ _{10} $ \cite{RN1563} and Cr single crystal \cite{RN1533}. Experiments of $\mu$SR, RIXS and NMR suggest presence of SDW in La$ _{3} $Ni$ _{2} $O$ _{7} $ or La$ _{3} $Ni$ _{2} $O$ _{7-\delta} $ \cite{2311.15717,2402.10485, 2401.12657, 2402.03952}. In particular, the transition temperature of SDW exhibits a distinct pressure-induced response compared to $T _\mathrm{CDW} $ \cite{2402.10485}. This highlights the need for further experimental investigations for relationship of CDW and SDW in this system.
	
	\textit{Summary}.---In this work, we report by performing comprehensive DFT calculations that the La$ _{3} $Ni$ _{2} $O$ _{7} $ possesses an antiferromagnetic ground state, with the strong FS nesting of the electronic flat band, which can drive the CDW transitions and trigger the MI transitions. In addition, we reveal that the competition between two different phases with apical oxygen vacancies can lead to a pressure-induced MI transition. These results are in good agreement with the experimental observations. The EPC in La$ _{3} $Ni$ _{2} $O$ _{7} $ is found insufficient to be responsible for superconductivity, suggesting that the Cooper pairing mechanism is unconventional, and may be from the AFM fluctuations. A complex phase diagram consistent with experiments are given. Our calculations explain well the physical origins behind a few experimental results, and give intensive comprehension for the physical properties such as superconductivity, CDW orders and roles of apical oxygen vacancies in La$ _{3} $Ni$ _{2} $O$ _{7} $, in which some results remain to be examined experimentally.

	\begin{acknowledgments}
		\textit{Acknowledgments}.--- The authors are indebted to Qing-Bo Yan, Wei Li, Jie Zhang, Xing-Zhou Qu, Dai-Wei Qu and Jia-Lin Chen for useful discussions. This work is supported in part by the National Key R$ \&$D Program of China (Grant No. 2018YFA0305800), the Strategic Priority Research Program of the Chinese Academy of Sciences (Grants No. XDB28000000), the National Natural Science Foundation of China (Grant No.11834014), and the Innovation Program for Quantum Science and Technology (No. 2021ZD0301800). B.G. is supported in part by the National Natural Science Foundation of China (Grant No. 12074378), the Chinese Academy of Sciences (Grants No. YSBR-030, No. JZHKYPT-2021-08, No. XDB33000000), Beijing Municipal Science and Technology Commission (Grant No. Z191100007219013).
	\end{acknowledgments}
	
%\bibliography{ref}	

	%%%%%%%%%%%%%%%%%%%%%%%%%%%%%%%%%%%%%%%%%%%%%%%%%%%%%%%%%%%%%%%%%%%%
	%======================================
	%========== Supplementary =============
	%======================================
	\newpage
	\clearpage
	\onecolumngrid
	\mbox{}
	\begin{center}
	{\large Supplementary Materials for} $\,$ \\	
	\bigskip	
	\textbf{\large{Antiferromagnetic Ground State, Charge Density Waves and Oxygen Vacancies Induced Metal-Insulator Transition in Pressurized La$ _{3} $Ni$ _{2} $O$ _{7} $}} \\		
	Yi \textit{et al}.
	\end{center}
\date{\today}
\setcounter{section}{0}
\setcounter{table}{0}
\setcounter{figure}{0}
\setcounter{equation}{0}
\renewcommand{\thetable}{S\arabic{table}}
\renewcommand{\theequation}{S\arabic{equation}}
\renewcommand{\thefigure}{S\arabic{figure}}
\setcounter{secnumdepth}{3}

	\section*{S1. Methods}
	Density functional theory (DFT) calculations were performed using two software packages: the Vienna ab initio simulation package (VASP) and QUANTUM ESPRESSO (QE) \cite{RN24, RN263}. In VASP, the atomic force convergence criterion was set to 1 meV/Å, with a plane-wave cutoff energy of 520 eV and a total energy convergence threshold of 10$^{-7}$ eV/atom. A $\Gamma$-centered Monkhorst-Pack k-mesh of 12×12×12 was utilized for self-consistent calculations of primative cell. In QE, the atomic force convergence criterion was set to 10$^{-7}$ Ry/a.u., with kinetic energy cutoffs for wavefunctions and charge density set to 100 Ry and 1200 Ry, respectively. The convergence threshold of total energy for wavefunction and ionic minimization was set at 10$^{-9}$ Ry. A $\Gamma$-centered Monkhorst-Pack k-mesh of 6×6×6 was used for self-consistent calculations, while a q-point grid of 3×3×3 was employed for electron-phonon coupling (EPC) calculations of the primative cell. The calculated lattice parameters and Wyckoff positions are listed in Table ~\ref{pristine}, consistent with previous experimental results \cite{RN1300}.
	
	The critical temperature ($\textit{T}_\textit{c}$) was determined using the McMillan semi-empirical formula \cite{RN128, RN1112}:
	\begin{equation}
		{\rm \textit{T}_\textit{c}}=\frac{\omega_{log}}{1.2}{\rm exp}\left[-\frac{1.04(1+\lambda)}{\lambda-\mu^*(1+0.62\lambda)}\right].\label{eq1}
	\end{equation}
	Here, $\mu^*$ is an empirical parameter describing Coulomb repulsion, and is set to $\mu^*$ = 0.10 for all calculations. $\omega_{log}$ and $\lambda$ were calculated by:
	\begin{equation}
		\omega_{log}=\exp\left[\frac{2}{\lambda}\int_0^\infty\frac{d\omega}{\omega}\alpha^2F(\omega)\log\omega\right],
	\end{equation}
	\begin{equation}
		\lambda(\omega)=2\int_0^\omega \frac{\alpha^2F(\omega)}{\omega}d\omega,
	\end{equation}
	where $\alpha^2F(\omega)$ denotes the Eliashberg electron-phonon spectral function.
	
	The Fermi surface (FS) nesting function is calculated by \cite{RN1480}
	\begin{equation}
		\xi(q)=\sum_{n n^{\prime}, \mathbf{k}} \delta\left(\varepsilon_{n,\mathbf{k}}-E_F\right)\delta\left(\varepsilon_{n^{\prime},\mathbf{k}+\mathbf{q}}-E_F\right), 
	\end{equation}
	where $\delta$ is the delta function, $\varepsilon_{n, \mathbf{k}}$ is the eigenvalue of band and $\mathbf{q}$ is the nesting vector.
	
	Phonon calculations and analysis of vibration modes are based on VASP and phonopy softwares \cite{RN24, RN22}.
	
	Harmonic phonon energy of different structures were calculated by implementing phonopy software \cite{RN22} with following equation:
	\begin{equation}
		E =\sum_{\mathbf{q} \nu} \hbar \omega(\mathbf{q} \nu) \left[\frac{1}{2}+\frac{1}{\exp (\hbar \omega(\mathbf{q}\nu) / k_{\textit{B}}T)-1}\right],
	\end{equation}
	where $ \nu $ is the band index and $ \mathbf{q} $ denotes the q-point.
	
	To calculate antiferromagnetic transition temperature $\textit{T}_\textit{N}$, we conducted Monte Carlo (MC) simulations on a 50$\times$50$\times$1 two-dimensional (2D) bilayer square lattice with periodic boundary conditions, with each temperature calculation containing $5\times 10^5 $ MC steps. We also performed MC simulations on the 30$\times$30$\times$30 three-dimensional (3D) square lattice. Our results indicate that due to the negligible \(J_3\), as seen in FIG. 1(a), the 3D and 2D double-layer square lattices yield the same $\textit{T}_\textit{N}$.
	
	\begin{table*}[!ht]
		%\begin{table*}\footnotesize\small\normalsize
		\renewcommand\arraystretch{1.15}
		\caption{Calculated lattice parameters a, b, and c (\AA) and Wyckoff sites of pristine phase for La$ _{3} $Ni$ _{2} $O$ _{7} $ at 0 and 30 GPa.}
		{\centering
			\begin{tabular}{lp{1.5cm}<{\centering}p{1.5cm}<{\centering}p{1.5cm}<{\centering}p{1.5cm}<{\centering}p{1.5cm}<{\centering}p{8cm}<{\centering}p{5cm}}
				\hline
				\hline
				&Pressure (GPa) &Space group &a &b &c &Wyckoff sites \\
				\hline
				&0 & $ \textit{Cmcm} $ &5.4057 &5.5303 &20.5526 & \begin{tabular}[c]{@{}l@{}}La: 8g(-0.250, 0.763, 0.319)\\ La: 4c(0.250, 0.248, 0.500)\\ Ni: 8g(0.250, 0.746, 0.404)\\ O: 8g(0.250, 0.802, 0.295)\\ O: 4c(0.250, 0.679, 0.500)\\ O: 8e(0.500, 0.000, 0.414)\\ O: 8e(0.000, 0.500, 0.389)\end{tabular}\\
				\\
				&30 & $ \textit{Fmmm} $ &5.1794 &5.1797 &19.5734 &\begin{tabular}[c]{@{}l@{}}La: 4b(0.000, 0.000, 0.500)\\ La: 8i(0.000, 0.000, 0.321)\\ Ni: 8i(0.000, 0.000, 0.098)\\ O: 4a(0.000, 0.000, 0.000)\\ O: 8i(0.000, 0.000, 0.204)\\ O: 16j(0.250, 0.250, 0.096)\end{tabular}\\
				\hline
				\hline	
		\end{tabular}}\label{pristine}
		\vspace{-0.2cm}
	\end{table*}
	
	\section*{S2. Magnetic calculations}
\begin{figure*}[!h]
		\centering
		\includegraphics[scale=0.5,angle=0]{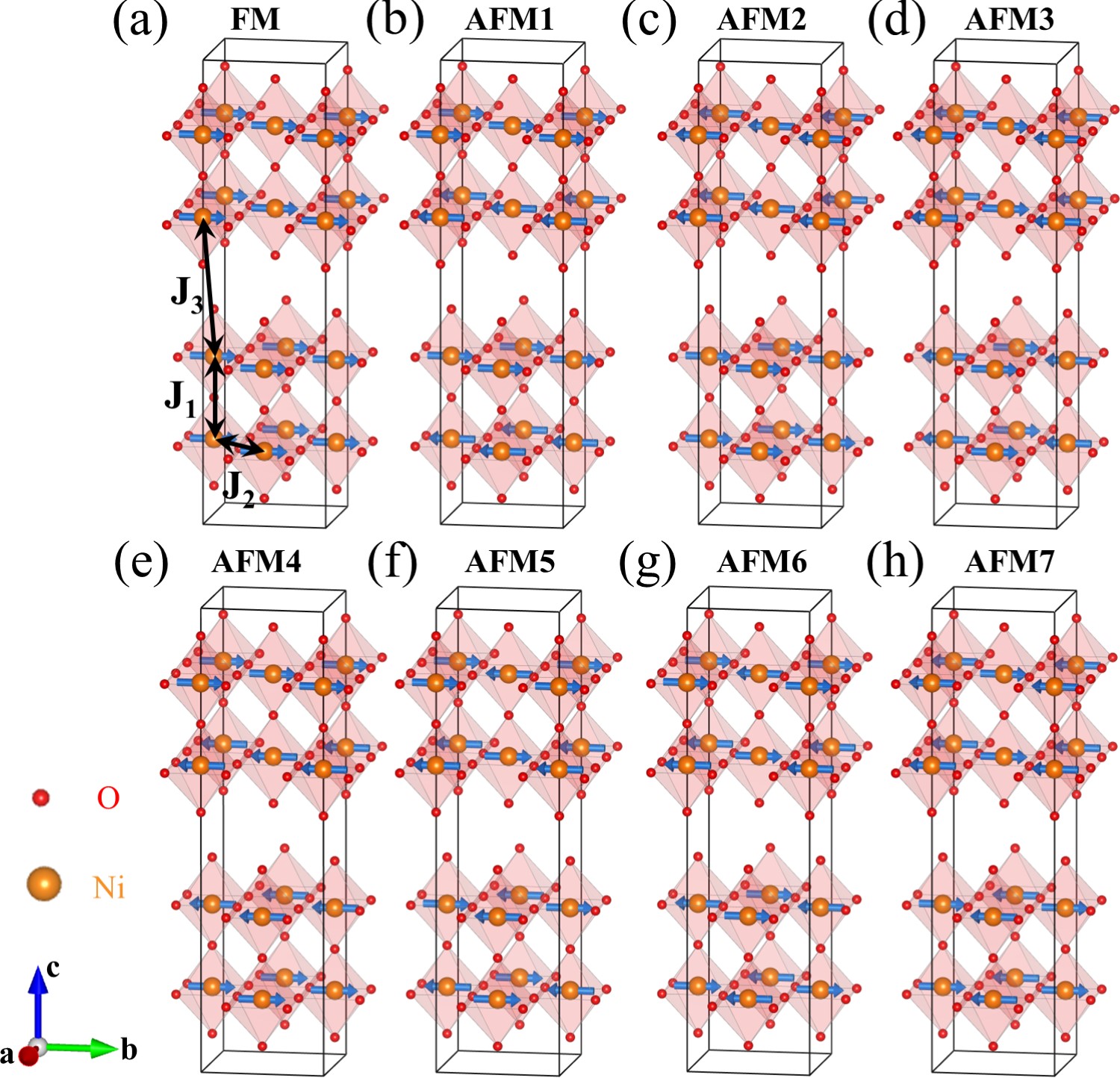}\\
		\caption{Calculated (a) ferromagnetic (FM) and (b-h) seven antiferromagnetic (AFM1-7) configurations by considering all possible combinations of $J_1$$s^2$, $J_2$$s^2$, and $J_3$$s^2$. Blue arrows indicate magnetic moments for different configurations.\\
			\\
		}\label{magconfigurations}
	\end{figure*}

\begin{figure*}[!h]
		\centering
		\includegraphics[scale=0.45,angle=0]{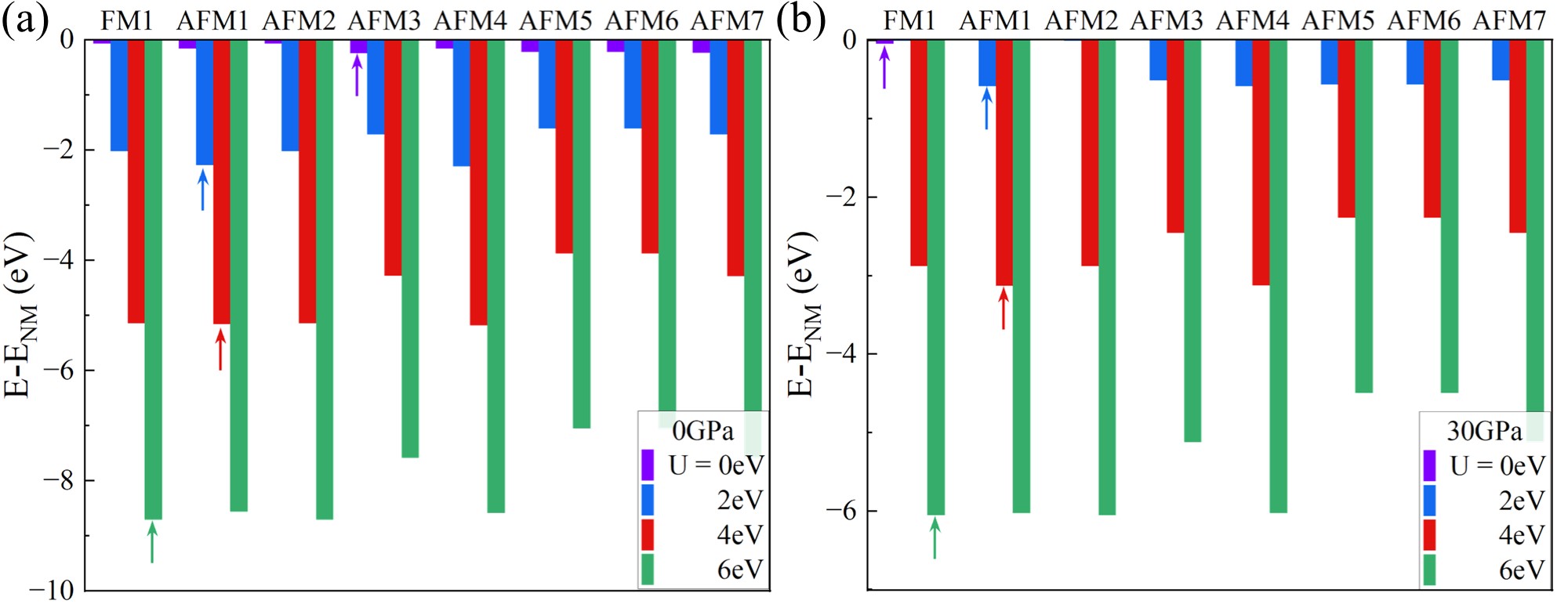}\\
		\caption{Energy of different magnetic configurations with respect to NM configuration with U= 0, 2, 4 and 6 eV for Ni-3d electrons at (a) 0 and (b) 30 GPa. The arrows indicate the magnetic ground state for each value of Hubbard U.
			\\
			\\
			\\
			\\
			\\
			\\}\label{magenergy}
	\end{figure*}

	All magnetic configurations, including one ferromagnetic and seven antiferromagnetic configurations, were initially tested with onsite Coulomb interactions U=0, 2, 4, and 6 eV for Ni-3d electrons as illustrated in FIG.~\ref{magconfigurations}. For U=0, all configurations exhibit similar energy to the nonmagnetic state, and the magnetic moments become zero under high pressure (see FIG.~\ref{magenergy}). Results for U=2 and 4 are almost the same, indicating that AFM1 is the magnetic ground state. As U increases to 6, the ground state transitions to FM configuration, possibly due to a double exchange mechanism as explained in a recent work \cite{RN1308}.
	
\begin{figure*}[!ht]
		\centering
		\includegraphics[scale=0.45,angle=0]{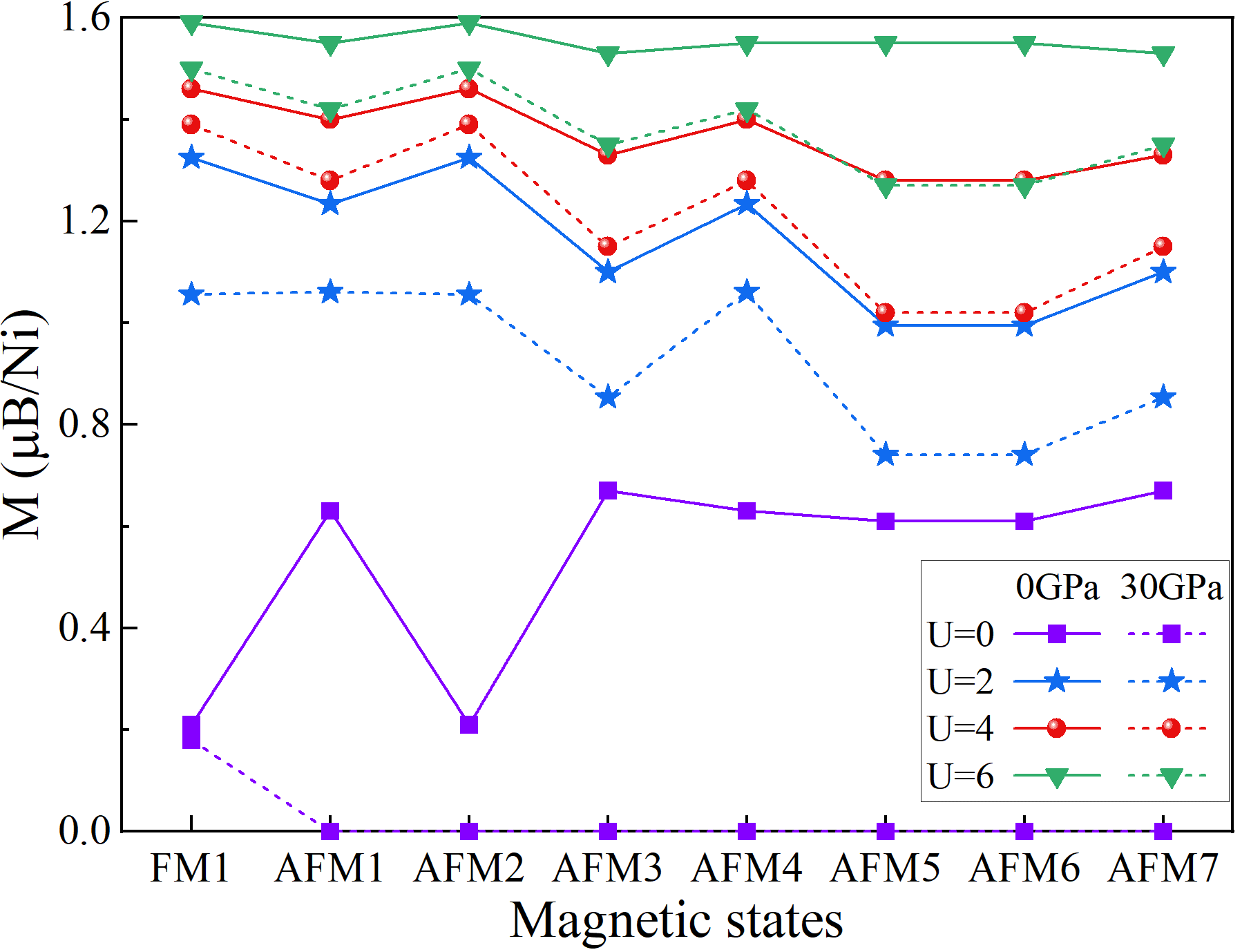}\\
		\caption{Magnetic moments of each Ni atom for different magnetic configurations with U= 0, 2, 4 and 6 eV for Ni-3d electrons.
			\\
			\\
			\\
		}\label{magneticmoment}
	\end{figure*}

	With increasing pressure, we found that the magnetic moments decrease by 0.1-0.6 $ \mu $B for all magnetic states in FIG.~\ref{magneticmoment}. Additionally, the relative energies of eight magnetic configurations with respect to the non-magnetic (NM) state decrease as seen in FIG. 1(b), indicating the pressure-induced suppression of magnetism.
	
	Recent DFT calculations in related studies have considered a reasonable value for Hubbard U in the range of 3-4 eV \cite{RN1300, RN1338, RN1324,2307.15276, 2307.07154, 2309.01148}. So we select U = 4 eV for analysis in our calculations.

\begin{figure*}[h]
		\centering
		\includegraphics[scale=1,angle=0]{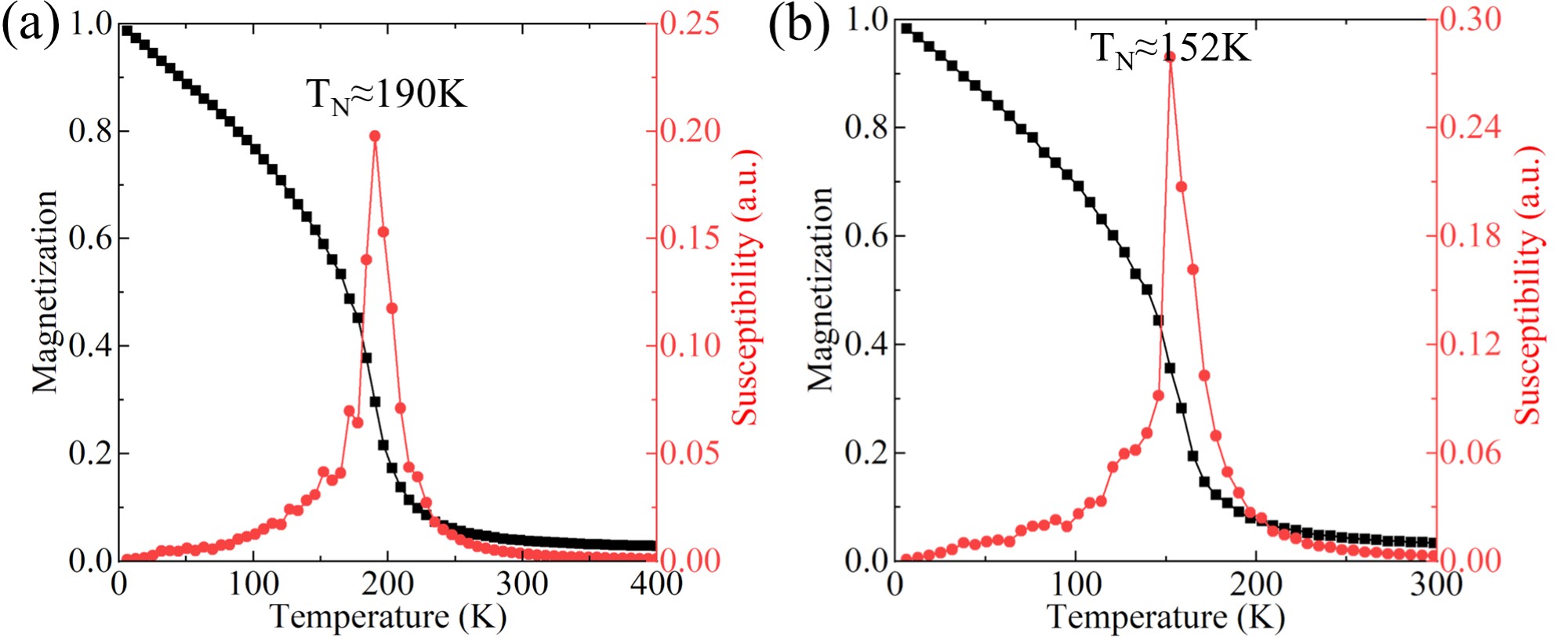}\\
		\caption{Magnetization and susceptibility of (a) CDW-Z1 phase at 0 GPa and (b) CDW-M1 phase at 30 GPa.\\
		}\label{MC}
	\end{figure*}

	The calculated $\textit{T}_\textit{N}$ and $J_1$$s^2$, $J_2$$s^2$, and $J_3$$s^2$ for different phases are listed in Table ~\ref{magtable}. $\textit{T}_\textit{N}$ are determined at the temperature corresponding to the discontinuity in susceptibility as illustrated in FIG.~\ref{MC}, where the magnetization is defined as the normalized average magnetic moment per magnetic site. It can be observed that the pristine phase exhibits a higher $\textit{T}_\textit{N}$ compared to the charge density wave (CDW) phase, while the HD phase with oxygen vacancies decreases $\textit{T}_\textit{N}$. The significant variation in $T_\textit{N, cal.}$ among different phases may explain the experimental discrepancy of $T_\textit{N, exp.}$.
	
	\begin{table*}[!ht]
		\renewcommand\arraystretch{1.15}
		\caption{$J_1$$s^2$, $J_2$$s^2$, $J_3$$s^2$ and $\textit{T}_\textit{N}$ for different phases.}
		\centering
		\begin{tabular}{lp{2.6cm}<{\centering}p{1.1cm}<{\centering}p{1.1cm}<{\centering}p{1.1cm}<{\centering}p{1.0cm}<{\centering}p{0.7cm}<{\centering}p{0.8cm}}
			\hline
			\hline
			&Phase &Pressure (GPa) &J$ _{1} $s$^{2} $ (meV) &J$ _{2} $s$^{2} $ (meV) &J$ _{3} $s$^{2} $ (meV) &T$_\textit{N}$ (K)\\
			\hline
			&Pristine-$ \mathit{Cmcm} $ &0 &-9.01 &41.8 &0 &393 \\ 
			&CDW-Z1 &0 &-28.0 &16.0 &0 &190 \\
			&La$ _{3} $Ni$ _{2} $O$ _{6.75} $ (HD) &0 &-19.7 &7.6 &0 &95 \\
			&Pristine-$ \mathit{Fmmm} $ &30 &-42.2 &25.7 &0 &304 \\ 
			&CDW-M1 &30 &-70.9 &11.3 &0 &152 \\
			\hline
			\hline	
		\end{tabular}
		\vspace{-0.2cm}
		\label{magtable}
	\end{table*}

	\newpage

	\section*{S3. Stable CDW phases}
\begin{figure*}[h]
		\centering
		\includegraphics[scale=0.5,angle=0]{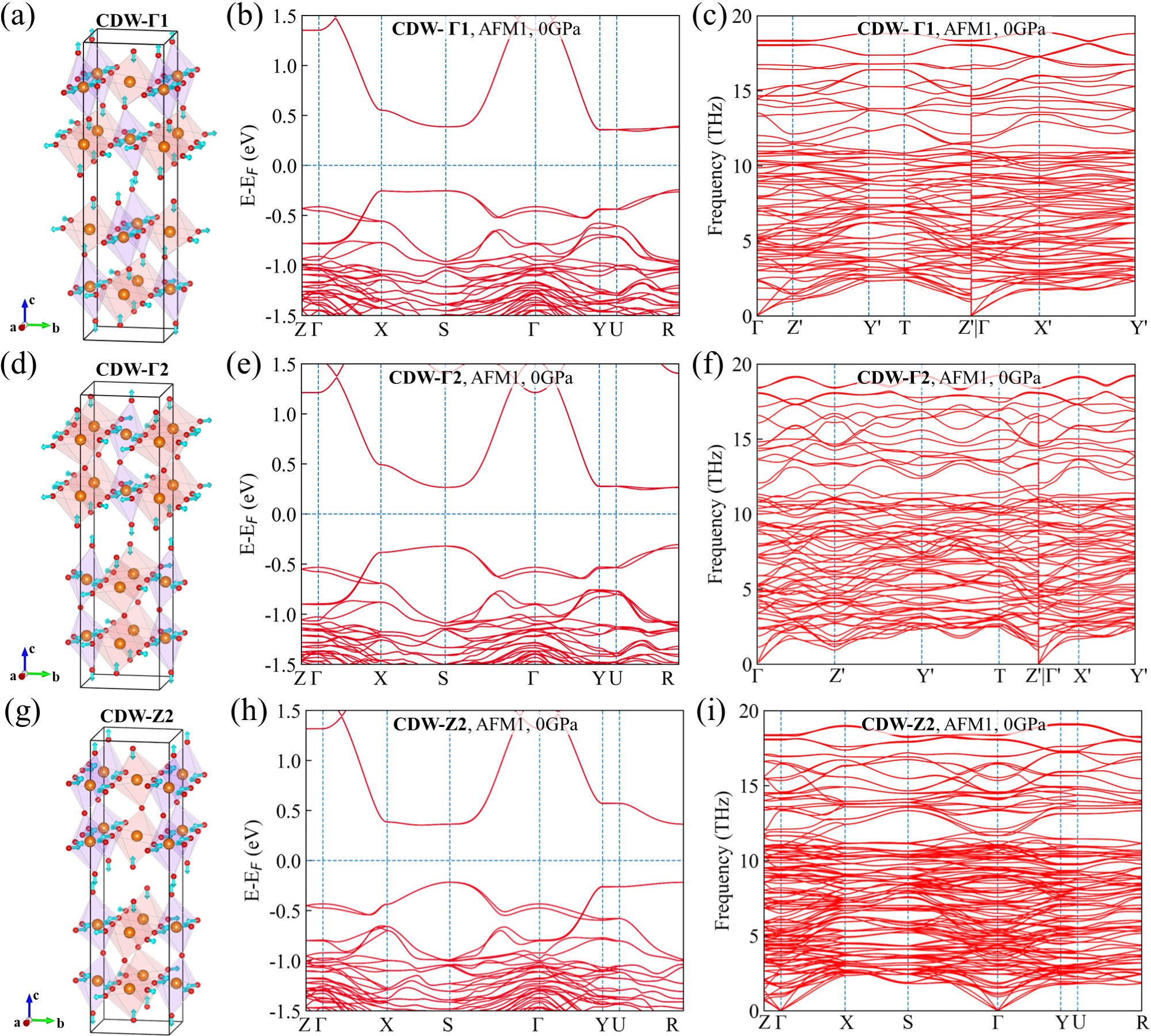}\\
		\caption{(a) Crystal structure, (b) band structure, and (c) phonon spectrum of CDW-$ \Gamma $1 phase corresponding to imaginary mode $ \Gamma $1 at 0 GPa. Oxygen octahedra with in-plane expansion (fat) and contraction (skim) are depicted with light red and purple colors, respectively. (d-f) The same as (a-c) but for CDW-$ \Gamma $2 phase corresponding to imaginary mode $ \Gamma $2 at 0 GPa. (g-i) The same as (a-c) but for CDW-Z2 phase corresponding to imaginary mode Z2 at 0 GPa. High symmetry points of these band structures and phonon spectra are shown in FIG. 1(e).
			\\
			\\
			\\}\label{G1}
		\vspace{1cm}
	\end{figure*}

	For the $ \textit{Cmcm} $ phase at low pressure, we uncovered that four imaginary modes lead to four stable CDW phases named CDW-$\Gamma1$, CDW-$\Gamma2$, CDW-Z1 and CDW-Z2, respectively. Their lattice parameters can be found in Table ~\ref{CDWlattice0}. Apart from the CDW-Z1 phase shown in FIG. 2 in the main text, the remaining three phases and their band structures and phonon spectra are shown in FIG.~\ref{G1}.
	
\begin{figure*}[!h]
		\centering
		\includegraphics[scale=0.6,angle=0]{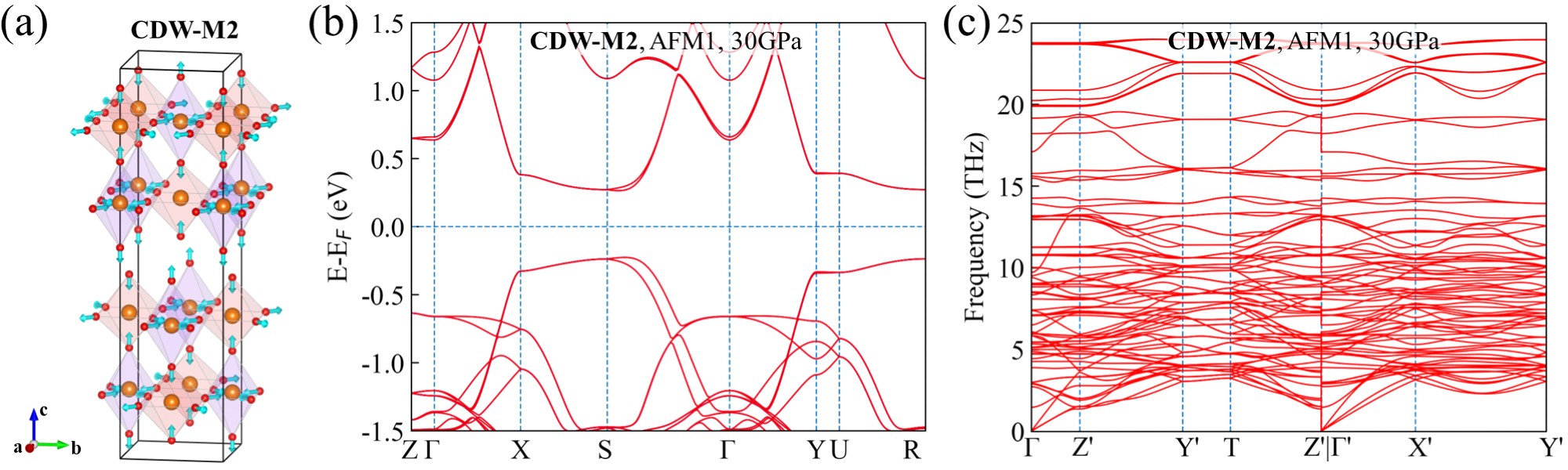}\\
		\caption{(a) Crystal structure, (b) band structure, and (c) phonon spectrum of CDW-M2 phase corresponding to imaginary mode M2 at 30 GPa. High symmetry points of (b) and (c) are shown in FIG. 1(e).}\label{M2}
		\vspace{1cm}
	\end{figure*}

	For the $ \textit{Fmmm} $ phase at high pressure, there are two imaginary modes that lead to two stable CDW phases named CDW-M1 and CDW-M2, respectively. Their lattice information can be found in Table ~\ref{CDWlattice30}. Except for CDW-M1 phase as shown in FIG. 2, the CDW-M2 phase and its band structure and phonon spectrum are presented in FIG.~\ref{M2}.
	
\begin{figure*}[!h]
		\centering
		\includegraphics[scale=0.3,angle=0]{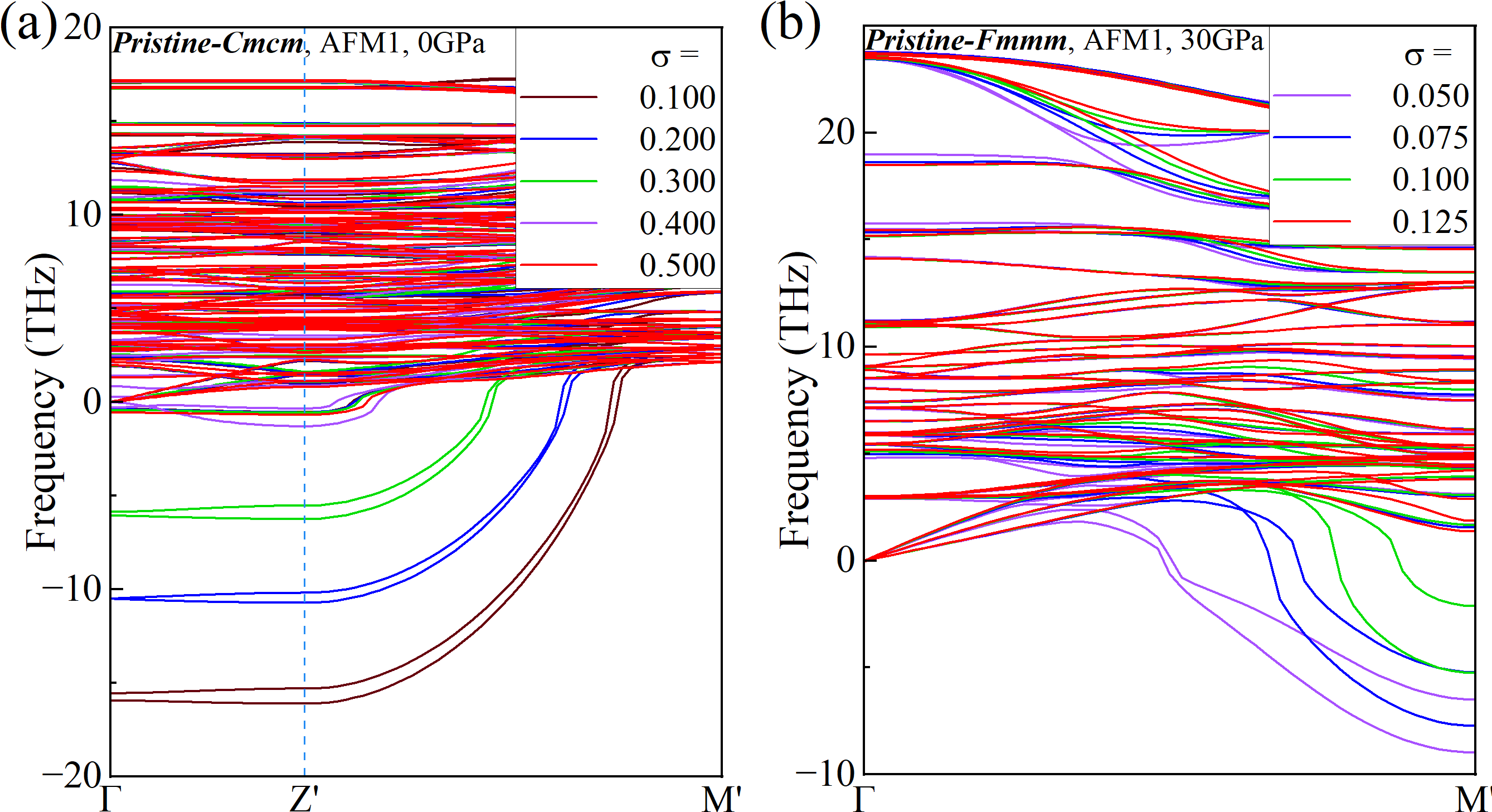}\\
		\caption{Phonon dispersions at (a) 0 and (b) 30 GPa for different \(\sigma\) values.}\label{sigma}
		\vspace{5cm}
	\end{figure*}

	The width of the smearing (\(\sigma\)) for partial occupancies set for each electronic orbital in our DFT calculation qualitatively represents the electronic temperature. Phonon dispersions at 0 and 30 GPa for different \(\sigma\) values are depicted in FIG.~\ref{sigma}, where the phonon soft modes near \(\Gamma\) and Z' at 0 GPa, as well as those near M' at 30 GPa, gradually harden until disappearing with increasing \(\sigma\). This indicates that temperature can suppress phonon softening, further confirming that the imaginary phonons at both low and high pressure originate from the CDW.
	
\begin{figure*}[!h]
		\centering
		\includegraphics[scale=0.6,angle=0]{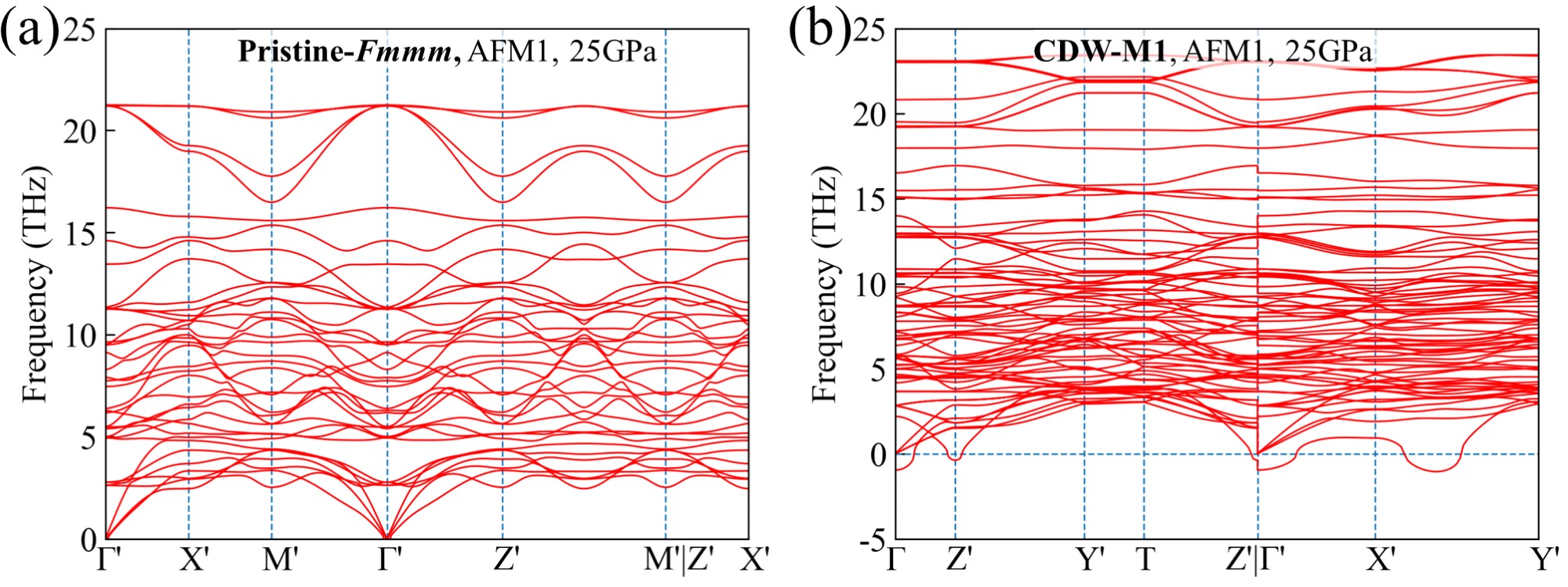}\\
		\caption{Phonon spectra of (a) pristine-Fmmm phase and (b) CDW-M1 phase corresponding to imaginary mode M1 at 25 GPa.}\label{20gpa}
		\vspace{4cm}
	\end{figure*}

	It is noteworthy that the CDW-Z1 phase consistently exhibits a stable phonon spectrum below 10 GPa. By comparing the harmonic phonon energies between the CDW-Z1 phase and the pristine $ \textit{Cmcm} $ phase, the estimated \textit{$\textit{T}_\textit{CDW}$} gradually decreases with increasing pressure, dropping to 0 K at about 7.5 GPa as shown in FIG. 5. On the other hand, the CDW-M1 phase only exhibits stable phonon spectra above 25 GPa. For instance, at 25 GPa, the phonon spectrum of CDW-M1 displays imaginary frequencies while the pristine $ \textit{Fmmm} $ phase is stable, as shown in FIG.~\ref{20gpa}. $\textit{T}_\textit{CDW}$ above 25 GPa increases with increasing pressure.
	
\begin{figure*}[!h]
		\centering
		\includegraphics[scale=0.33,angle=0]{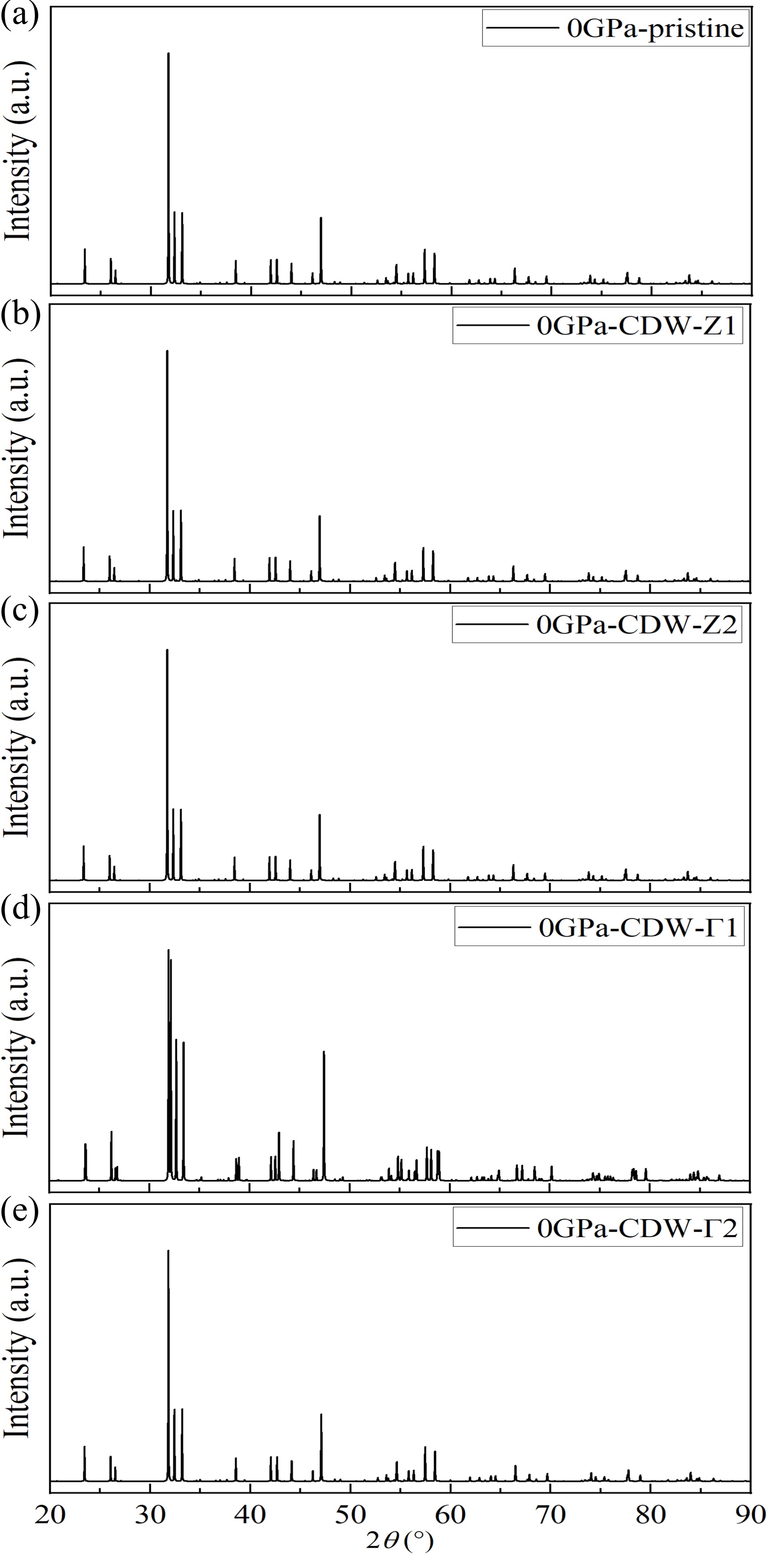}\\
		\caption{Simulated XRD patterns of (a) pristine-Cmcm, (b) CDW-Z1, (c) CDW-Z2, (d) CDW-$ \Gamma $2 and (e) CDW-$ \Gamma $2 phases at 0 GPa. }\label{xrd0gpa}
		\vspace{6cm}
	\end{figure*}

\begin{figure*}[!ht]
		\centering
		\includegraphics[scale=0.33,angle=0]{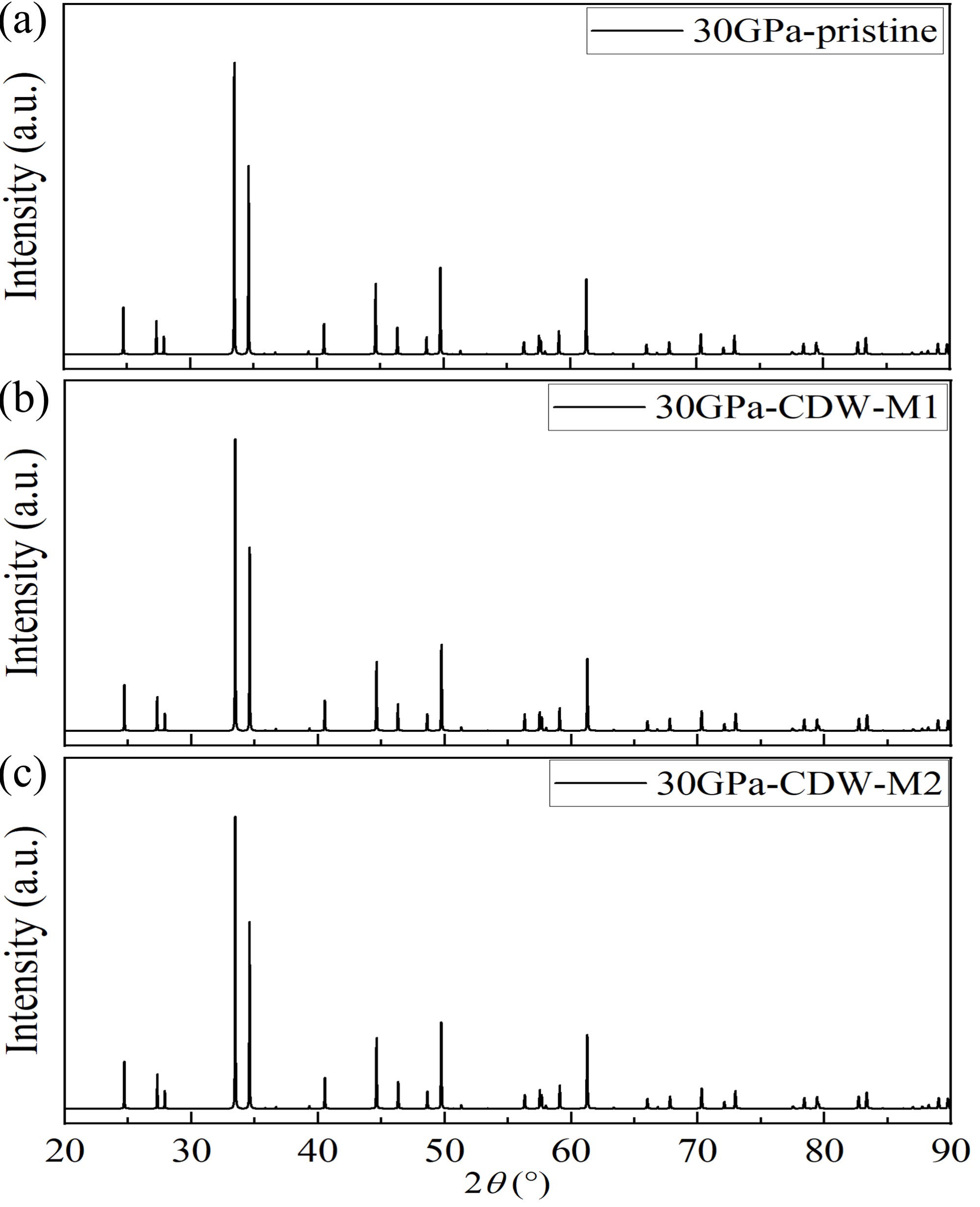}\\
		\caption{Simulated XRD patterns of (a) pristine-Fmmm, (b) CDW-M1 and (c) CDW-M2 phases at 30 GPa. }\label{xrd30gpa}
	\end{figure*}

\begin{figure*}[!ht]
		\centering
		\includegraphics[scale=0.6,angle=0]{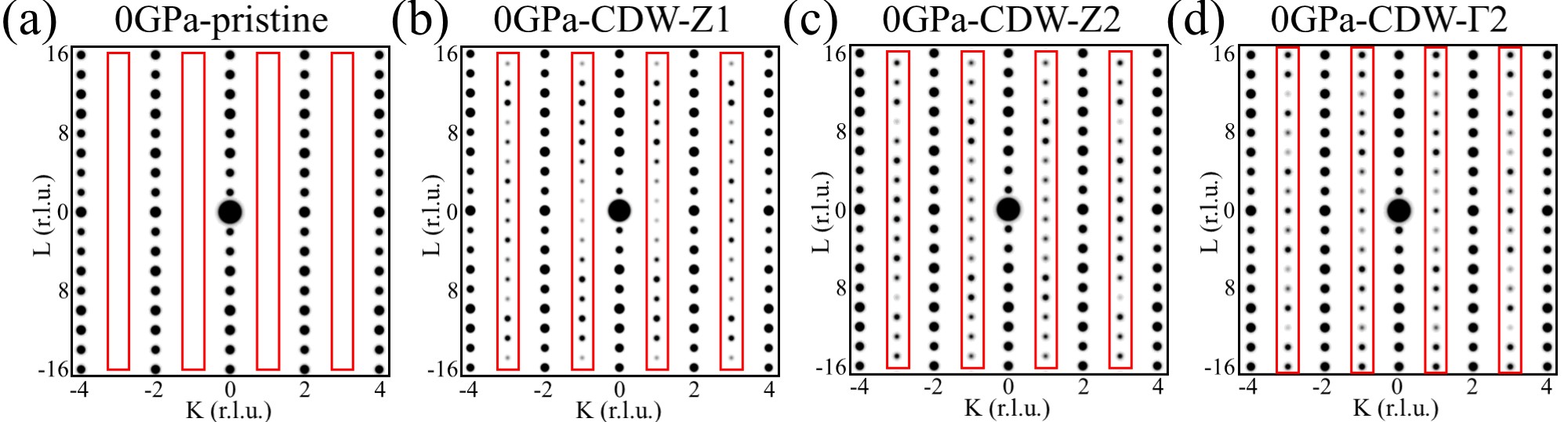}\\
		\caption{Simulated diffraction patterns of (a) pristine-Cmcm, (b) CDW-Z1, (c) CDW-Z2 and (d) CDW-$ \Gamma $2 phases along the [100] zone axis at 0 GPa. Miller indices $ \textit{K} $ and $ \textit{L} $ are in reciprocal lattice unit (r.l.u.). Emergent Bragg peaks of CDW phase compared with pristine phase are highlighted by red boxes.}\label{pattern0}
	\end{figure*}

\begin{figure*}[!ht]
		\centering
		\includegraphics[scale=0.6,angle=0]{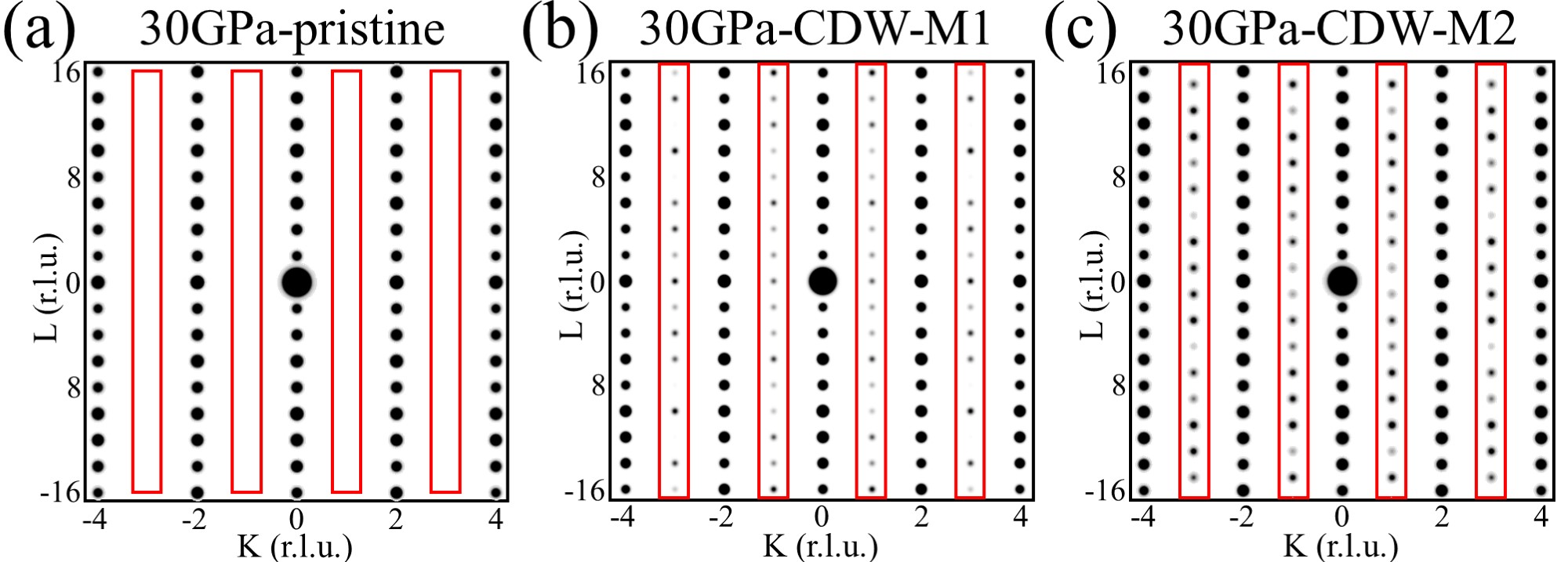}\\
		\caption{Simulated diffraction patterns of (a) pristine-Fmmm, (b) CDW-M1 and (c) CDW-M2 phases along the [100] zone axis at 30 GPa. Miller indices $ \textit{K} $ and $ \textit{L} $ are in reciprocal lattice unit (r.l.u.). Emergent Bragg peaks of CDW phase compared with pristine phase are highlighted by red boxes.}\label{pattern30}
	\end{figure*}

	As seen in FIG.~\ref{xrd0gpa} and ~\ref{xrd30gpa}, most CDW phases exhibit almost the same peak positions and intensities comparing to the pristine phase at both 0 and 30 GPa in XRD spectra, except CDW-$\Gamma$1 with a distinctive Bravais lattice. Diffraction patterns of CDW phases show emergent spots with odd Miller indices $ \textit{K} $ with pristine phases in FIG.~\ref{pattern0} and ~\ref{pattern30}.
	
	\begin{table*}[!ht]
		\caption{Calculated lattice parameters $ \textit{a} $, $ \textit{b} $, and $ \textit{c} $ (\AA) and Wyckoff sites of CDW phases for La$ _{3} $Ni$ _{2} $O$ _{7} $ at 0 GPa.}
		{\centering
			\begin{tabular}{lp{1.5cm}<{\centering}p{1.5cm}<{\centering}p{1.5cm}<{\centering}p{1.5cm}<{\centering}p{1.5cm}<{\centering}p{1.5cm}<{\centering}p{5cm}<{\centering}p{5cm}}
				\hline
				\hline
				&Pressure (GPa) &Phase &Space group &$ \textit{a} $ &$ \textit{b} $ &$ \textit{c} $ &Wyckoff sites \\
				\hline
				&0 &CDW-$\Gamma$1 &$\textit{P}\mathit{2}_{1}\textit{/m}$ &10.6243 &5.4031 &5.5305 & \begin{tabular}[c]{@{}l@{}}
					La: 2e(0.999, 0.750, 0.247)\\ La: 2e(0.361, 0.250, 0.945)\\
					La: 2e(0.638, 0.250, 0.581)\\ Ni: 2e(0.191, 0.750, 0.844)\\
					Ni: 2e(0.807, 0.750, 0.646)\\ O: 4f(0.171, 0.505, 0.081)\\
					O: 4f(0.222, 0.995, 0.620)\\ O: 2e(0.414, 0.750, 0.016)\\
					O: 2e(0.593, 0.750, 0.594)\\ O: 2e(0.997, 0.750, 0.677)\\ \end{tabular}\\
				\\
				
				&0 &CDW-$\Gamma$2 &$ \textit{Amm2} $ &5.3971 &20.5989 &5.5288 & \begin{tabular}[c]{@{}l@{}}
					La: 4e(-0.500, 0.681, -0.734)\\ La: 4d(0.000, 0.320, -0.260)\\
					La: 2b(-0.500, 0.500, -0.754)\\ La: 2a(0.000, 0.500, -0.259)\\
					Ni: 4e(-0.500, 0.404, -0.246)\\ Ni: 4d(0.000, 0.596, -0.755) \\
					O: 8f(-0.754, 0.914, -0.009)\\ O: 8f(-0.246, 0.389, -0.992)\\
					O: 4e(-0.500, 0.297, -0.297)\\ O: 4d(0.000, 0.207, 0.190)\\
					O: 2b(-0.500, 0.500, -0.186)\\ O: 2a(0.000, 0.500, -0.828)\\
				\end{tabular}\\
				\\
				&0 &CDW-Z1 &$ \textit{Pnma} $ &5.5368 &5.3987 &20.5764 & \begin{tabular}[c]{@{}l@{}}
					La: 4c(0.252, 0.250, 0.499)\\ La: 4c(0.725, 0.750, 0.681)\\
					La: 4c(0.248, 0.750, 0.180)\\ Ni: 4c(0.746, 0.250, 0.596)\\
					Ni: 4c(0.263, 0.250, 0.097)\\ O: 8d(0.999, 0.504, 0.086)\\
					O: 8d(0.982, 0.003, 0.611)\\ O: 4c(0.212, 0.250, 0.203)\\
					O: 4c(0.681, 0.250, 0.707)\\ O: 4c(0.822, 0.250, 0.498)\\
				\end{tabular}\\
				\\
				&0 &CDW-Z2 &$ \textit{Pmmn} $ &5.4006 &20.6027 &5.5241 & \begin{tabular}[c]{@{}l@{}}
					La: 2b(0.000, 0.500, -0.488)\\ La: 2a(0.000, 0.000, -0.006)\\
					La: 4e(0.500, 0.320, -0.011)\\ La: 4e(0.500, 0.819, -0.515)\\
					Ni: 4e(0.000, 0.404, -0.991)\\ Ni: 4e(0.000, 0.904, -0.500)\\
					O: 8g(0.255, 0.914, -0.762)\\ O: 8g(0.755, 0.389, -0.057)\\
					O: 2b(0.000, 0.500, -0.919)\\ O: 4e(0.000, 0.797, -0.549)\\
					O: 4e(0.000, 0.293, -0.057)\\ O: 2a(0.000, 0.000, -0.439)\\
				\end{tabular}\\
				\\
				
				\hline
				\hline	
				\\
				\\
				\\
				\\
				\\
				\\
				\\
				\\
		\end{tabular}}\label{CDWlattice0}
	\end{table*}
	
	\begin{table*}[!ht]
		\caption{Calculated lattice parameters $ \textit{a} $, $ \textit{b} $, and $ \textit{c} $ (\AA) and Wyckoff sites of CDW phases for La$ _{3} $Ni$ _{2} $O$ _{7} $ at 30 GPa.}
		{\centering
			\begin{tabular}{lp{1.5cm}<{\centering}p{1.5cm}<{\centering}p{1.5cm}<{\centering}p{1.5cm}<{\centering}p{1.5cm}<{\centering}p{1.5cm}<{\centering}p{5cm}<{\centering}p{5cm}}
				\hline
				\hline
				&Pressure (GPa) &Phase &Space group &$ \textit{a} $ &$ \textit{b} $ &$ \textit{c} $ &Wyckoff sites \\
				\hline
				&30 &CDW-M1 &$ \textit{Cmmm} $ &5.179 &19.585 &5.178 & \begin{tabular}[c]{@{}l@{}}
					La: 4j(0.000, 0.179, -0.500)\\ La: 4i(0.500, 0.180, 0.000)\\
					La: 2d(0.000, 0.000, -0.500)\\ La: 2b(0.500, 0.000, 0.000)\\
					Ni: 4j(0.000, 0.598, -0.500)\\ Ni: 4i(0.000, 0.098, 0.000)\\
					O: 16r(0.754, 0.404, -0.253)\\ O: 2a(0.000, 0.000, 0.000)\\
					O: 2c(0.500, 0.000, -0.500)\\ O: 4j(0.000, 0.295, -0.500)\\
					O: 4i(0.500, 0.297, 0.000)\\
				\end{tabular}\\
				\\
				
				&30 &CDW-M2 & $ \textit{Cmcm} $ &5.1794 &19.5770 &5.1808 & \begin{tabular}[c]{@{}l@{}}
					La: 4c(0.500, 0.929, -0.250)\\ La: 4c(0.500, 0.750, -0.250)\\
					La: 4c(0.000, 0.930, -0.750)\\ Ni: 4c(0.500, 0.652, -0.750)\\ Ni: 4c(0.000, 0.652, -0.250)\\ O: 16h(0.753, 0.654, -0.496)\\ O: 4c(0.500, 0.749, -0.750)\\ O: 4c(0.500, 0.547, -0.750)\\ O: 4c(0.000, 0.545, -0.250)\\
				\end{tabular}\\
				\\
				\hline
				\hline
				\\
				\\
				\\
				\\
				\\
				\\	
				\\
				\\
				\\
				\\	
		\end{tabular}}\label{CDWlattice30}
	\end{table*}
	
	\section*{S4. Role of Oxygen vacancies}
	
\begin{figure*}[h]
		\centering
		\includegraphics[scale=0.5,angle=0]{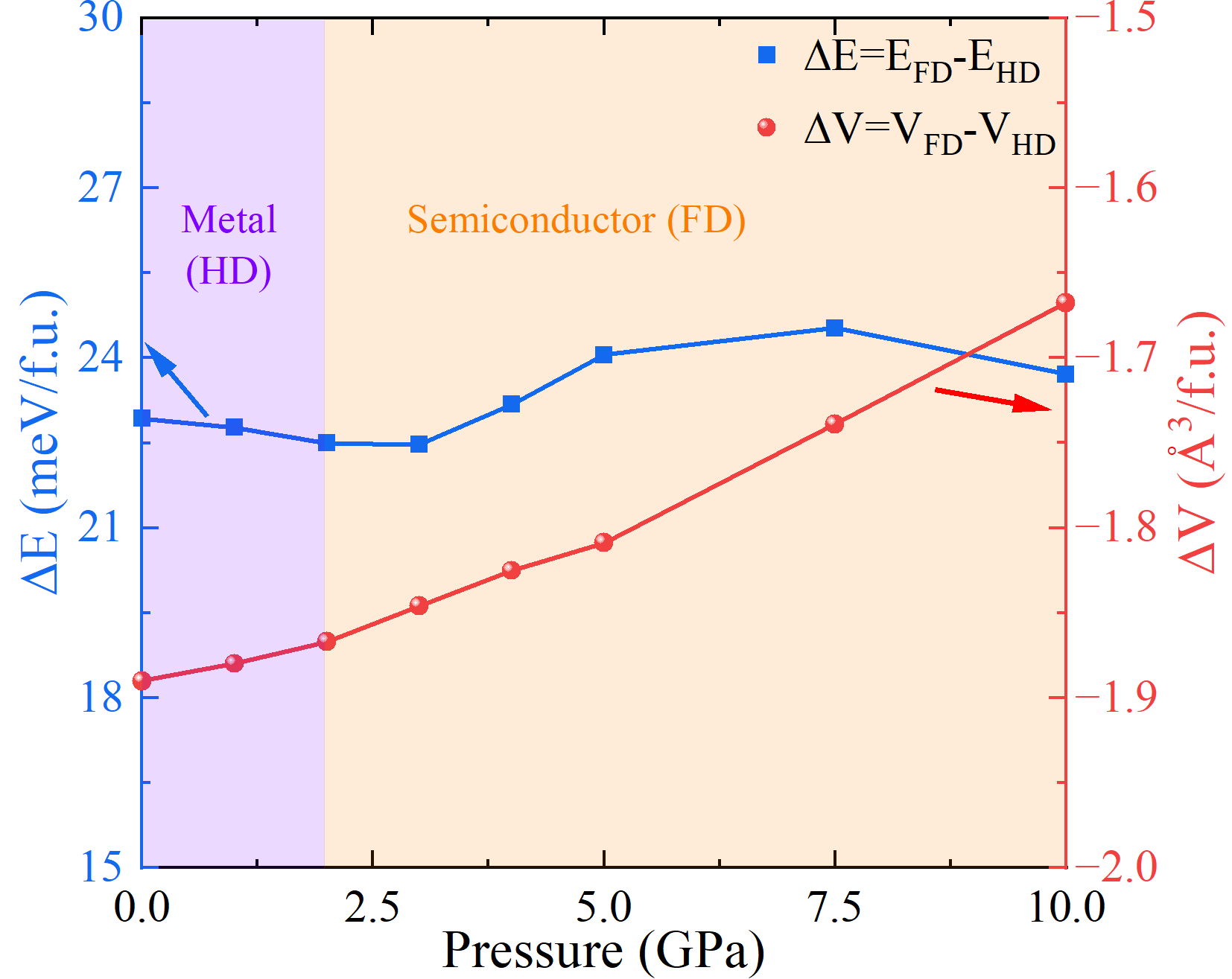}\\
		\caption{(a) Energy difference ($\Delta$E) and (b) volume difference ($\Delta$V) between the HD and FD phases under pressure.}\label{oxygenvacanies}
	\end{figure*}

	As concluded in the main text, with apical oxygen vacancies, there appear two different phases at low and high pressures owing to pressure dependent enthalpy ($H$). The half distortion (HD) phase has a lower $H$ than the full distortion (FD) phase below 2 GPa, while the FD phase has a lower $H$ above 2 GPa. The $H$ can be divided into two components, $H=E+PV$, where $E$, $P$, and $V$ represent energy, pressure, and volume, respectively. By comparing $E$ and $V$ between the HD and FD phases under pressure in FIG.~\ref{oxygenvacanies}, although the HD phase consistently exhibits lower $E$ at 0-10GPa, the smaller $V$ of the FD phase makes it thermodynamically more stable under high pressure.
	
\end{document}